\documentclass[10pt]{article}

\usepackage{epsfig}
\usepackage{amsfonts}
\usepackage{color}
\usepackage{amssymb}

\newcommand{\eg}{\emph{e.g.\/}}
\newcommand{\s}{\sqsubseteq}
\newcommand{\N}{\mathcal{N}}
\newcommand{\ES}{\mathcal{S}}

\newcommand{\hato}{\mathcal{H}_o}
\newcommand{\epe}{\E_{\preceq}}
\newcommand{\E}{\mathcal{E}}
\newcommand{\V}{\mathcal{V}}
\newcommand{\ACCA}{\mathcal{H}}
\newcommand{\ans}{\mathit{ans}}
\newcommand{\indo}{\mathit{ind_S(o)}}
\newcommand{\PI}{\mathcal{P}}
\newcommand{\obj}{\mathit{Obj}}
\newcommand{\EL}{\mathcal{L}}
\newcommand{\ok}{\hfill $\Box$}
\newcommand{\ie}{\emph{i.e.\/}}
\newcommand{\qe}{\textsc{Qe}}
\newcommand{\qidone}{\emph{QID}$_1$}
\newcommand{\subp}{\mathit{SP}}

\long\def\comment#1{}

\newcommand{\qid}{\emph{QID}}
\newcommand{\qpone}{\emph{QP}$_1$}
\newcommand{\qp}{\emph{QP}}
\newcommand{\query}{\textsc{Query}}
\newcommand{\procask}{\textsc{Ask}}
\newcommand{\proctell}{\textsc{Tell}}
\newcommand{\procaqs}{\textsc{Ask-Queue-Server}}
\newcommand{\proctqs}{\textsc{Tell-Queue-Server}}
\newcommand{\ask}{\texttt{ask}}
\newcommand{\tell}{\texttt{tell}}

\newcommand{\liq}{\textsc{Query-Cache}}

\newcommand{\orq}{\textsc{Output-Request-Queue}}

\newcommand{\aq}{\textsc{Answer-Queue}}
\newcommand{\askq}{\textsc{Ask-Queue}}
\newcommand{\tellq}{\textsc{Tell-Queue}}

\newcommand{\clo}{\texttt{closed}}

\newcommand{\pri}{\texttt{principal}}
\newcommand{\dec}{\texttt{declined}}
\newcommand{\tot}{\texttt{total}}
\newcommand{\prt}{\texttt{partial}}
\newcommand{\id}{Query-ID}

\newcommand{\bl}
  {\begin{list}{--}{
    \setlength{\topsep}{4pt}
    \setlength{\itemsep}{0pt}
    \setlength{\parsep}{4pt}
    \setlength{\partopsep}{0pt}}}
\newcommand{\el}{\end{list}}

\newtheorem{definition}{Definition}
\newtheorem{proposition}{Proposition}
\newtheorem{lemma}{Lemma}

\newtheorem{corollary}{Corollary}
\begin{document}

\title{Query processing in distributed, taxonomy-based information sources}

\author{Carlo Meghini\thanks{Consiglio Nazionale delle Ricerche, Istituto della
    Scienza e delle Tecnologie della Informazione, Pisa, Italy} \and Yannis
  Tzitzikas\thanks{Department of Computer Science, University of Crete,
    Heraklion, Greece and Institute of Computer Science, Foundation for Research
    and Technology -- Hellas} \and Veronica Coltella\thanks{Consiglio Nazionale
    delle Ricerche, Istituto della Scienza e delle Tecnologie della
    Informazione, Pisa, Italy} \and Anastasia Analyti\thanks{Institute of
    Computer Science, Foundation for Research and Technology -- Hellas, Crete,
    Greece} }

\date{}

\maketitle

\begin{abstract}
  We address the problem of answering queries over a distributed
  information system, storing objects indexed by terms organized
  in a taxonomy. The taxonomy consists of subsumption
  relationships between negation-free DNF formulas on terms and
  negation-free conjunctions of terms.  In the first part of the
  paper, we consider the centralized case, deriving a
  hypergraph-based algorithm that is efficient in data
  complexity.  In the second part of the paper, we consider the
  distributed case, presenting alternative ways implementing the
  centralized algorithm. These ways descend from two basic
  criteria: direct \emph{vs.} query re-writing evaluation, and
  centralized \emph{vs.}  distributed data or taxonomy
  allocation. Combinations of these criteria allow to cover a
  wide spectrum of architectures, ranging from client-server to
  peer-to-peer. We evaluate the performance of the various
  architectures by simulation on a network with $O(10^4)$ nodes,
  and derive final results. An extensive review of the relevant
  literature is finally included.
\end{abstract}


\section{Introduction}
\label{chap:intro}

Consider an information source $\ES$ structured as a tetrad
$\ES=(T, \preceq, \obj, I),$ where $T$ is a set of terms,
$\preceq$ is a taxonomy over concepts expressed using $T$ (e.g.
$\mathtt{(Animal \wedge FlyingObject}) \vee \mathtt{Penguin}
\preceq \mathtt{Bird}$), $\obj$ is a set of objects and $I$ is
the interpretation, that is a function from $T$ to $\PI(\obj)$,
assigning an extension (\ie, a set of objects) to each term.  Now
assume that there is a set $\N$ of such sources $\N=\{\ES_1,
\ldots, \ES_n\},$ all sharing the same set of objects $\obj$ and
related by taxonomic relationships amongst concepts of different
sources. These relationships are called \emph{articulations} and
aim at bridging the inevitable naming, granularity and contextual
heterogeneities that may exist between the taxonomies of the
sources (for some examples see \cite{TzitzikasMeghiniER03}).  For
example, the taxonomy of a source $\ES_1$ could be the following:
$\{~ \mathtt{Penguin} \preceq \mathtt{Animal}$, $\mathtt{Pelican}
\preceq \mathtt{Animal}$, $\mathtt{Ostrich} \preceq
\mathtt{Animal}$, $\mathtt{(Animal \wedge FlyingObject}) \vee
\mathtt{Penguin} \vee \mathtt{Ostrich} \preceq \mathtt{Bird}
~\}$.  $\ES_1$ could have an articulation to a source $\ES_2$
like $\{~ \Pi\iota\nu\gamma\kappa o\upsilon \acute{\iota}\nu
o\varsigma \preceq \mathtt{Penguin}$, $
\Pi\epsilon\lambda\epsilon\kappa\acute{\alpha} \nu o\varsigma
\preceq \mathtt{Pelican} ~\}$, an articulation to a source
$\ES_3$ like $\{~ \mathtt{Animale} \wedge \mathtt{Alato} \preceq
\mathtt{Birds} ~\}$, and an articulation to two sources $\ES_4,
\ES_5$ of the form: $\{~ (\mathtt{Fliegentier}) \vee
(\mathtt{Animal} \wedge \mathtt{Volant}) \preceq (\mathtt{Animal}
\wedge \mathtt{FlyingObject}) ~\}$.

Network of sources of this kind are nowadays commonplace. For instance, the
objects may be web pages, and a source may be a portal serving a specific
community endowed with a vocabulary used for indexing web pages. The objects may
be library resources such as books, serials, or reports, and a source may be a
library describing the content of the resources according to a local
vocabulary. The objects may be a category of commercial items, such as cars, and
a source may be an e-commerce site which sells the items. And so
on. Articulations may be drawn from language dictionaries, or may be the result
of cooperation agreements, such as in the case of sources belonging to the same
organization. In certain cases, articulations can be constructed automatically,
for instance using the data-driven method proposed in
\cite{TzitzikasMeghiniCIA03}. Moreover, sources and articulations expressed in a
syntactically richer language, such as a Semantic web language, are typically
mapped down to the sources and articulations we assume, for computational
reasons~\cite{rousset2}.

In this paper we address the problem of answering Boolean queries over networks
of this kind of sources. The work is carried out in three stages.

First, the theoretical aspects of query evaluation against a source are
analyzed, and an algorithm is derived which extends a hypergraph-based method
for satisfiability of propositional Horn clauses. The algorithm is conceptually
very simple and has polynomial time complexity with respect to the size of
$\obj.$ This is in fact the theoretical lower bound.

Secondly, we derive different implementations of the query evaluation algorithm,
all of them exploiting the use of a cache for optimization reasons. The
different implementations stem from the considerations of two orthogonal
criteria: evaluation mode and data allocation. The first criterion leads to two
alternative approaches: direct query evaluation \emph{vs.} query re-writing. The
second criterion leads to four possibilities, corresponding to the centralized
\emph{vs.} the distributed allocation of the taxonomy or of the interpretation.
When considered in combinations, these two criteria give rise to five
interesting distributed architectures, ranging from the well-known client/server
architecture to the recently proposed peer-to-peer (P2P) systems.

In considering the data allocation criterion, we have made the assumption that
the sources are willing to make (some of) their data available for storing in a
centralized way. This may not always be the case. If it is not the case, then
only one implementation is possible, namely the one based on the P2P
architecture, which reflects the model of a source at the physical level.

Thirdly, the derived implementations are evaluated from a performance point of
view, in terms of response time. The performance evaluation has been carried out
by simulating the implementations on a network of 11400 sources. The network
has been configured based on the parameters derived in a study on the Gnutella
network. The results of the simulations show that the client-server
implementation is, perhaps not surprisingly, the one offering the shortest
response time. Amongst the other architectures, the best performance is
attained, perhaps surprisingly, by centralizing the taxonomy while keeping the
interpretations distributed and executing query evaluation in two stages. This
is due to the fact that re-writing the query avoids multiple accesses to the
same source, but this gain can be appreciated only if the taxonomy is
centralized, so that a single access (to the taxonomy server) is necessary to
perform the query re-writing.

In sum, the three main results of the paper are: (i) a query evaluation
procedure for a source; (ii) five different optimized implementations of the
query evaluation procedure, corresponding to the considered architectures;
all implementations are optimized, in that they make use of a cache; and (iii) a
ranking of these algorithms, based on their performance.

The paper is structured as follows: Section~\ref{sec:found} introduces the model
of information system studied in the paper, formulates the query evaluation
problem, and derives a sound and complete query evaluation procedure for the
centralized case. Section~\ref{sec:ba} illustrates a basic method for carrying
out query evaluation, putting the theoretical notions developed in the previous
Section into a concrete software perspective. Section~\ref{sec:appr} discusses
possible ways of implementing the basic method in a distributed setting,
deriving five significant architectures. For each architecture, a description of
the behaviour of the involved components is provided. Section~\ref{sec:perfev}
presents an evaluation of the performance of the five architectures in terms of
response time for query evaluation. Section \ref{sec:rw} compares our work with
related work and Section \ref{sec:conc} concludes the paper.

Finally, we would like to mention that initial ideas of this work
have appeared in our conference
paper~\cite{ArticulatedODBASE04}. This work
extends~\cite{ArticulatedODBASE04} by providing (i) the query
evaluation principles in Section 3, (ii) the algorithms and
architectures for network query evaluation in Section 4, and
(iii) the performance evaluation in Section 5.

\section{Foundations}
\label{sec:found}

This Section defines information sources and the query evaluation
problem. The algorithmic foundations of this problem are given
and an efficient query evaluation method is provided. These
results will be applied later, upon studying networks of sources.

\subsection{The model}
\label{sec:em}

The basic notion of the model is that of \emph{terminology:} a terminology $T$
is a non-empty set of terms. From a terminology, \emph{queries} can be
defined.

\begin{definition}[Query] \label{def:QueryLang} \emph{The \emph{query language}
    associated to a terminology $T,$ $\EL_T,$ is the language defined by the
    following grammar, where $t$ is a term of $T:$
    \begin{tabbing}
      ind \= m \= ::= \= \kill
      \> $q$ \> ::= \> $d ~ | ~ q \vee d$ \\
      \> $d$ \> ::= \> $t ~ | ~ t \wedge d.$
    \end{tabbing}
    An instance of $q$ is called a \emph{query,} while an instance of $d$ is
    called a \emph{conjunctive query} and a {\em disjunct} of $q$ whenever $d$
    occurs in $q.$} \ok
\end{definition}

Terms and conjunctive queries can be used for defining taxonomies.

\begin{definition}[Taxonomy] \label{def:TaxDef} 
  \emph{A \emph{taxonomy} over a terminology $T$ is a pair $(T,\preceq)$ where
    $\preceq$ is any set of pairs $(q, d)$ where $q$ is any query and $d$ is a
    conjunctive query.} \ok
\end{definition}

For example, if $T=\{a1, a2, b1, b2, b3, c1\}$ then a taxonomy over $T$ could be
$(T,\preceq)$ where (using an infix notation) $\{(b1 \wedge b2)\vee
b3~\preceq~a1\wedge a2, ~a1\wedge a2~\preceq~c1\}$.

If $(q,q') \in \; \preceq,$ we say that $q$ is subsumed by $q'$ and we write
$q\preceq q'.$

\begin{definition}[Interpretation] \label{def:Int} \emph{An
    \emph{interpretation} for a terminology $T$ is a pair $(\obj,I)$, where
    $\obj$ is a finite, non-empty set of objects and $I$ is a total function
    assigning a possibly empty set of objects to each term in $T,$ \ie~$I: T
    \rightarrow \PI(\obj).$ } \ok
\end{definition}

Interpretations are used to define the semantics of the query language:

\begin{definition}[Query extension] \label{def:QExt} 
  \emph{Given an interpretation $I$ of a terminology $T$ and a query $q\in
    \EL_T,$ the \emph{extension of q in I,} $q^I,$ is defined as follows:
    \begin{enumerate}
    \item $(q\vee d)^I=~q^I\cup~ d^I$
    \item $(d\wedge t)^I=~d^I\cap~ t^I$
    \item $t^I=I(t).$ \ok
    \end{enumerate}}
\end{definition}

Since $\cdot^I$ is an extension of the interpretation function $I,$ we will
simplify notation and will write $I(q)$ in place of $q^I.$ We can now define an
\emph{information source} (or simply \emph{source}).

\begin{definition}[Information source] \label{def:IS} 
    \emph{An \emph{information source} $S$ is a 4-tuple
    $S=(T_S,\preceq_S,\obj_S,I_S)$, where $(T_S,\preceq_S)$ is a taxonomy and
    $(\obj_S,I_S)$ is an interpretation for $T_S.$ } \ok
\end{definition}

When no ambiguity will arise, we will omit the subscript in the components of
sources and equate $I$ with $(\obj,I),$ for simplicity. Moreover, given a source
$S=(T, \preceq,\obj,I)$ and an object $o\in\obj,$ the \emph{index of o in S,}
$\indo,$ is given by the terms in whose interpretation $o$ belongs, \ie:
\[
\indo=\{t\in T~|~o\in I(t)\}.
\]
\indent The interpretations that reflect the semantics of subsumption are as
customary called \emph{models,} defined next.

\begin{definition}[Models of a source] \label{def:Models}
  \emph{Given two interpretations $I$, $I'$ of the same terminology $T,$
    \begin{enumerate}
    \item $I$ is a {\em model} of the taxonomy $(T,\preceq)$ if $q \preceq q'$
      implies $I(q) \subseteq I(q');$ \label{item:int}
    \item $I$ is smaller than $I',$ $I \leq I'$, if $I(t) \subseteq I'(t)$ for
      each term $t \in T;$
    \item $I$ is a {\em model} of a source $S=(T,\preceq,\obj,I')$ if it is a
      model of $(T,\preceq)$ and $I'\leq I.$ \ok
    \end{enumerate}
    }
\end{definition}

Based on the notion of model, the answer to a query is finally defined.

\begin{definition}[Answer] \label{def:AnswerQS}
  \emph{Given a source $S=(T,\preceq,\obj,I)$ and a query $q\in\EL_T,$ the
    \emph{answer of q in S,} $\ans(q,S),$ is given by $\ans(q,S)=\{o\in
    \obj~|~o\in J(q) \mbox{ for all models $J$ of $S$}\}.$ } \ok
\end{definition}

Since we are exclusively interested in query evaluation, we can restrict
ourselves to simpler notions of sources and queries, which are equivalent to
those defined so far from the answer point of view. To begin with, we observe
that a pair $(q,q')$ in a taxonomy is interpreted (in
Definition~\ref{def:Models} point~\ref{item:int}) as an implication
$q\rightarrow q'.$ Now, by a simple truth table argument, it can be easily
verified that the propositional formula:
\[
(C_1\vee\ldots\vee C_n) \rightarrow (t_1\wedge\ldots\wedge t_m)
\]
where each $C_i$ is any propositional formula, is logically equivalent to the
formula:
\[
(C_1 \rightarrow t_1) \wedge (C_1 \rightarrow t_2) \wedge \ldots \wedge (C_1
\rightarrow t_m) \wedge \ldots \wedge (C_n \rightarrow t_1) \wedge (C_n
\rightarrow t_2) \wedge \ldots \wedge (C_n \rightarrow t_m),
\]
in that the two formulae have the same models. Based on this equivalence, the
\emph{simplification} of a taxonomy $(T,\preceq)$ is defined as
the taxonomy $(T,\preceq^s),$ where:
\[
\preceq^s=\{(C,t)~|~(C_1\vee\ldots\vee C_n)~\preceq~(t_1\wedge\ldots\wedge t_m),
~C\in\{C_1,\ldots,C_n\}, ~t\in\{t_1,\ldots,t_m\}\}.
\]
Correspondingly, the simplification of a source $S=(T,\preceq,\obj,I)$ is
defined to be the source $S^s=(T,\preceq^s,\obj,I).$ It is not difficult to see
that:
\begin{proposition} \label{prop:red}
  \emph{$J$ is a model of a source $S$ if and only if it is a model of $S^s.$ }
  \ok
\end{proposition}
The simplification of the taxonomy in the previous example is given by: \\
$\{(b1 \wedge b2)~\preceq^s~a1,~(b1 \wedge b2)~\preceq^s~a2, ~b3~\preceq^s~a1,
~b3~\preceq^s~a2, ~b3~\preceq^s~a2, ~a1\wedge a2~\preceq^s~c1\}.$

For simplicity, from now on $\preceq$ and $S$ will stand for $\preceq^s$
and $S^s,$ respectively.

Finally, non-term queries can be replaced by term queries by inserting
appropriate relationships into the taxonomy. Specifically:

\begin{proposition} \label{prop:termqonly}
  \emph{For all sources $S=(T,\preceq,\obj,I)$ and non-term queries
    $q\in\EL_T,$ let $t_q$ be any term not in $T$ and moreover
    \begin{eqnarray*}
      T^q & = & T\cup\{t_q\} \\
      \preceq^q & = & \preceq \cup~\{(t_1\wedge\ldots\wedge t_m,t_q)
      |~t_1\wedge\ldots\wedge t_m \mbox{ is a disjunct of }q\} \\
      I^q & = & I \cup \{(t_q,\emptyset)\}.
    \end{eqnarray*}
    Then, $\ans(q,S)=\ans(t_q,S^q)$ where $S^q=(T^q, \preceq^q,\obj,I^q).$ }
  \ok
\end{proposition}
In practice, the terminology $T^q$ includes one additional term $t_q,$ which has
an empty interpretation and subsumes each query disjunct $t_1\wedge\ldots\wedge
t_m.$ The size of $S^q$ is clearly polynomial in the size of $S$ and $q.$

In light of the last Proposition, the problem of query evaluation amounts to
determine $\ans(t,S)$ for given term $t$ and source $S,$ while the corresponding
decision problem consists in checking whether $o\in\ans(t,S),$ for a given
object $o.$

\subsection{The decision problem}
\label{sec:dec}

Given a source $S=(T,\preceq,\obj,I)$, $o \in \obj$, and $t \in T$, the decision
problem $o \in \ans(t,S)$ has an equivalent formulation in propositional
datalog. We define the propositional datalog program $P_{S}$ as follows:
\[
P_S = C_S \cup I_S \cup Q_S
\]
where
\begin{eqnarray*}
  C_S & = & \{t'\leftarrow t_1,\ldots,t_m~|~(t_1\wedge\ldots\wedge
  t_m,t')\in\preceq^r \} \\
  I_S & = & \{u\leftarrow ~|~u\in\indo\} \\
  Q_S & = & \{\leftarrow t\}
\end{eqnarray*}

It is easy to see that:

\begin{lemma}\label{lemma:exts}
  \emph{For all sources $S=(T,\preceq,\obj,I),$ $o\in\obj$ and $t\in T,$
    $o\in\ans(t,S)$ iff $P_S$ is unsatisfiable. } \ok
\end{lemma}

Based on Lemma \ref{lemma:exts}, the decision problem $o \in \ans(t,S)$ is connected to directed
B-hypergraphs, which are introduced next. We will mainly use definitions and
results from~\cite{gallo93}.

A \emph{directed hypergraph} is a pair $\ACCA=(\V,\E),$ where $\V=\{v_1,
v_2,\ldots, v_n\}$ is the set of vertexes and $\E=\{E_1, E_2,\ldots, E_m\}$ is
the set of directed hyperedges, where $E_i=(\tau(E_i),\chi(E_i))$ with
$\tau(E_i),\chi(E_i)\subseteq \V$ for $1\leq i\leq m.$ $\tau(E_i)$ is said to be
the \emph{tail} of $E_i,$ while $\chi(E_i)$ is said to be the \emph{head} of
$E_i.$ A \emph{directed B-hypergraph} (or simply \emph{B-graph}) is a directed
hypergraph, where the head of each hyperedge $E_i,$ denoted as $h(E_i),$ is a
single vertex.

\begin{figure}[htbp]
  \centering
  \begin{minipage}{7cm}
    \begin{picture}(5.5,2.5)
      \put(1,2){\makebox(0.5,0.5){$a1$}}
      \put(0.5,1){\makebox(0.5,0.5){$a2$}}
      \put(1.5,1){\makebox(0.5,0.5){$a3$}}
      \put(1,0){\makebox(1,0.5){$b1\wedge b2$}}
      \put(3.25,0){\makebox(1,0.5){$b1\wedge b3$}}
      \put(0,0){\makebox(0.5,0.5){$b3$}}
      \put(3,2){\makebox(0.5,0.5){$b1$}}
      \put(4.75,2){\makebox(0.5,0.5){$b2$}}
      \put(2.5,1){\makebox(0.5,0.5){$c1$}}
      \put(3.5,1){\makebox(0.5,0.5){$c2$}}
      \put(4.5,1){\makebox(1,0.5){$c2\wedge c3$}}

      \multiput(0.75,1.5)(2,0){2}{\vector(1,2){0.25}}
      \multiput(1.75,1.5)(2,0){2}{\vector(-1,2){0.25}}
      \multiput(3.75,0.5)(1.25,1){2}{\vector(0,1){0.5}}
      \put(0.25,0.5){\vector(1,2){0.25}}
      \put(1.5,0.5){\vector(-1,1){0.5}}
    \end{picture}
  \end{minipage}
  \begin{minipage}{5.75cm}
    \begin{picture}(5.75,3.5)
      \put(0,1.5){\makebox(0.5,0.5){$a1$}}
      \put(1.5,1.5){\makebox(0.5,0.5){$a2$}}
      \put(3,1.5){\makebox(0.5,0.5){$b2$}}
      \put(4.5,1.5){\makebox(0.5,0.5){$c2$}}
      \put(1.5,2.5){\makebox(0.5,0.5){$a3$}}
      \put(3,2.5){\makebox(0.5,0.5){$b3$}}
      \put(4.5,2.5){\makebox(0.5,0.5){$c3$}}
      \put(3,0.5){\makebox(0.5,0.5){$b1$}}
      \put(4.5,0.5){\makebox(0.5,0.5){$c1$}}
      \multiput(1.5,2.5)(1.5,0){2}{\vector(-2,-1){1}}
      \multiput(1.5,1.75)(1.5,0){3}{\vector(-1,0){1}}
      \put(3,1){\line(-1,2){0.37}}
      \put(4.5,1.5){\vector(-2,-1){1}}
      \put(4.5,2.5){\line(-1,-2){0.38}}
      \put(4.5,0.75){\vector(-1,0){1}}
      \put(4.25,3){\oval(2,.75)[t]}
      \put(5.25,3){\line(0,-1){1.25}}
      \put(5.25,1.75){\vector(-1,0){.25}}
      \put(4.5,0.5){\oval(2.5,.75)[b]}
      \put(5.25,1.25){\oval(1,1)[tr]}
      \put(5.75,0.5){\line(0,1){.75}}
    \end{picture}
  \end{minipage}
\caption{A taxonomy and its B-graph}
  \label{fig:extsource}
\end{figure}
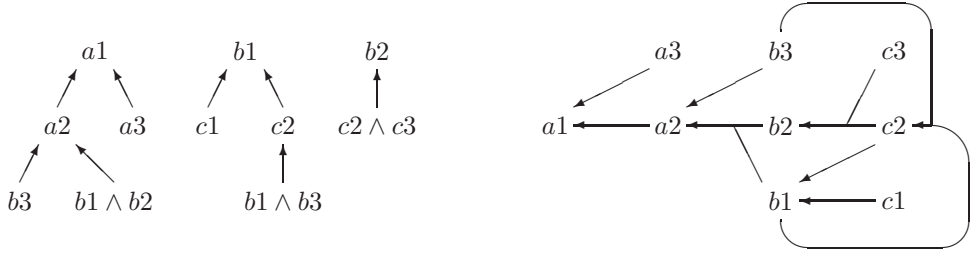

A taxonomy can naturally be represented as a B-graph whose hyperedges represent
one-to-one the subsumption relationships of the transitive reduction of the
taxonomy.  In particular, the \emph{taxonomy B-graph} of a taxonomy
$(T,\preceq)$ is the B-graph $\ACCA=(T,\epe),$ where
\[
\epe = \{(\{t_1,\ldots,t_m\},u)~|~(t_1\wedge\ldots\wedge t_m,u)\in\preceq^r\}.
\]
Figure~\ref{fig:extsource} left presents a taxonomy, whose B-graph is shown in
the same Figure right.

A \emph{path} $P_{st}$ of length $q$ in a B-graph $\ACCA=(\V,\E)$ is a sequence
of nodes and hyperedges
\[
P_{st}=(s=v_1,E_{i_1},v_2,E_{i_2},\ldots,E_{i_q},v_{q+1}=t),
\]
where: $s\in\tau(E_{i_1}),$ $h(E_{i_q})=t$ and
$h(E_{i_{j-1}})=v_j\in\tau(E_{i_j})$ for $2\leq j\leq q.$ If $P_{st}$ exists,
$t$ is said to be \emph{connected} to $s.$ If $t\in\tau(E_{i_1}),$ $P_{st}$ is
said to be a \emph{cycle}; if all hyperedges in $P_{st}$ are distinct, $P_{st}$
is said to be \emph{simple.} A simple path is \emph{elementary} if all its
vertexes are distinct.

A \emph{B-path} $\pi_{st}$ in a B-graph $\ACCA=(\V,\E)$ is a minimal (with respect to
deletion of vertexes and hyperedges) hypergraph $\ACCA_\pi=(\V_\pi, \E_\pi),$
such that:
\begin{enumerate}
\item $\E_\pi\subseteq\E$
\item $\{s,t\}\subseteq\V_\pi$
\item $x\in\V_\pi$ and $x\neq s$ imply that $x$ is connected to $s$ in
  $\ACCA_\pi$ by means of a cycle-free simple path.
\end{enumerate}
Vertex $y$ is said to be \emph{B-connected} to vertex $x$ if a B-path $\pi_{xy}$
exists in $\ACCA.$

B-graphs and satisfiability of propositional Horn clauses are strictly
related. The B-graph {\em associated to} a set of Horn clauses has 3 types of
directed hyperedges to represent each clause:

\begin{itemize}
\item the clause $p \leftarrow q_1\wedge q_2\wedge \ldots\wedge q_s$ is
  represented by the hyperedge $(\{q_1,q_2, \ldots, q_s\}, p);$
\item the clause $ \leftarrow q_1\wedge q_2\wedge \ldots\wedge q_s$ is
  represented by the hyperedge $(\{q_1,q_2, \ldots, q_s\}, \mathit{false});$
\item the clause $p \leftarrow $ is represented by the hyperedge
  $(\{\mathit{true}\}, p).$
\end{itemize}
The following result is well-known:
\begin{proposition}[\cite{gallo93}] \label{prop:sat}
  \emph{A set of propositional Horn clauses is satisfiable if and only if in the
    associated B-graph, \emph{false} is not B-connected to \emph{true}.  } \ok
\end{proposition}

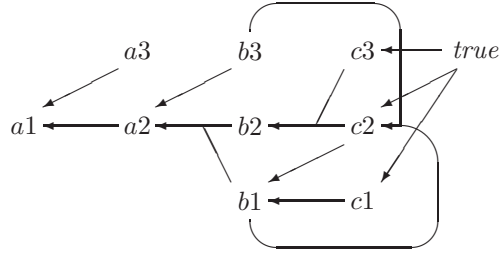
\begin{figure}[t]
  \centering
  \begin{picture}(6.5,3.5)

    \put(0,1.5){\makebox(0.5,0.5){$a1$}}
    \put(1.5,1.5){\makebox(0.5,0.5){$a2$}}
    \put(3,1.5){\makebox(0.5,0.5){$b2$}}
    \put(4.5,1.5){\makebox(0.5,0.5){$c2$}}
    \put(1.5,2.5){\makebox(0.5,0.5){$a3$}}
    \put(3,2.5){\makebox(0.5,0.5){$b3$}}
    \put(4.5,2.5){\makebox(0.5,0.5){$c3$}}
    \put(3,0.5){\makebox(0.5,0.5){$b1$}}
    \put(4.5,0.5){\makebox(0.5,0.5){$c1$}}
    \put(6,2.5){\makebox(0.5,0.5){\emph{true}}}


    \multiput(1.5,2.5)(1.5,0){2}{\vector(-2,-1){1}}
    \multiput(1.5,1.75)(1.5,0){3}{\vector(-1,0){1}}
    \put(3,1){\line(-1,2){0.37}}
    \put(4.5,1.5){\vector(-2,-1){1}}
    \put(4.5,2.5){\line(-1,-2){0.38}}
    \put(4.5,0.75){\vector(-1,0){1}}
    \put(5.8,2.75){\vector(-1,0){0.8}}
    \put(6,2.5){\vector(-2,-1){1}}
    \put(6,2.5){\vector(-2,-3){1}}


    \put(4.25,3){\oval(2,.75)[t]}
    \put(5.25,3){\line(0,-1){1.25}}
    \put(5.25,1.75){\vector(-1,0){.25}}
    \put(4.5,0.5){\oval(2.5,.75)[b]}
    \put(5.25,1.25){\oval(1,1)[tr]}
    \put(5.75,0.5){\line(0,1){.75}}

  \end{picture}
  \caption{An object graph}
  \label{fig:accaaqo}
\end{figure}

We now proceed to show the role played by B-connection in query evaluation. For
a source $S=(T, \preceq,\obj,I)$ and an object $o\in\obj,$ the \emph{object
  decision graph} (simply the \emph{object graph}) is the B-graph
$\ACCA_o=(T,\E_o),$ where
\[
\E_o = \epe \cup \bigcup\{(\{\mathit{true}\},u)~|~u\in\indo\}.
\]
Figure~\ref{fig:accaaqo} presents the object graph for the taxonomy shown in
Figure~\ref{fig:extsource} and an object $o$ such that $\indo=\{c1,c2,c3\}.$

We can now prove:

\begin{proposition} \label{prop:poqees}
  \em{For all sources $S=(T,\preceq,\obj,I),$ terms $t\in T,$ and objects
    $o\in\obj,$ $o\in\ans(t,S)$ iff \emph{t} is B-connected to
    \emph{true} in the object graph $\hato.$ \\
    {\em Proof}: From Lemma~\ref{lemma:exts}, $o\in\ans(t,S)$ iff $P_S$ is
    unsatisfiable iff (by Proposition~\ref{prop:sat}) \emph{false} is
    B-connected to \emph{true} in the associated B-graph. By construction,
    $\hato$ is the B-graph associated to $P_S,$ where \emph{t} plays the role
    of \emph{false.} \ok}
\end{proposition}

\subsection{An algorithm for query evaluation}
\label{sec:psqees}

A typical approach for query evaluation is resolution, recently studied for
peer-to-peer networks~\cite{rousset1,rousset2,adjman06}.  Here, we propose a
simpler method to perform query evaluation, based on B-graphs. Our method relies
on the following result, which is just a re-phrasing of
Proposition~\ref{prop:poqees}:

\begin{corollary} \label{prop:iteli}
  \emph{For all sources $S=(T,\preceq,\obj,I),$ $o\in\obj$ and term queries
    $t\in T,$ $o\in\ans(t,S)$ if and only if either $o\in I(t)$ or there exists
    a hyperedge $(\{u_1,\ldots,u_r\},t)\in\epe$ such that $o\in\bigcap
    \{\ans(u_i,S)~|~1\leq i\leq r\}.$} \ok
\end{corollary}
This corollary simply ``breaks down'' Proposition~\ref{prop:poqees} based on the
distance between $t$ and \emph{true} in the object graph $\hato.$ If $o\in
I(t),$ then $t\in\indo,$ hence there is a hyperedge (in fact, a simple arc) from
\emph{true} to $t$ in $\hato,$ which are 1 hyperedge distant from each other. If
$o\not\in I(t),$ then there are at least two hyperedges in between \emph{true}
and $t.$ Let us assume that $h$ is the one whose head is $t.$ Since $t$ is
B-connected to \emph{true,} each term $u_i$ in the tail of $h$ is B-connected to
\emph{true.} But this simply means, again by Proposition~\ref{prop:poqees}, that
$o\in\ans(u_i,S)$ for all the terms $u_i,$ and so we have the forward direction
of the Corollary. The backward direction of the Corollary is straightforward.
Notice that, by point 3 in the definition of B-path, $t$ is connected to each
$u_i$ by a cycle-free simple path; this fact is used by the procedure \qe~in
order to correctly terminate in presence of loops in the taxonomy B-graph
$\ACCA.$

\begin{figure}
{\small
\begin{tabbing}
Le \= x \= x \= x \= x \= x \= x \= x \= x \= \kill
\qe($x:$ \textbf{term}~; $A:$ \textbf{set of terms}); \\
1. \> $R$ $\leftarrow$ $I(x)$ \\
2. \> \textbf{for each} hyperedge $\langle \{u_1,\ldots,u_r\},x\rangle$ in
$\ACCA$ \textbf{do} \\
3. \> \> \textbf{if} $\{u_1,\ldots,u_r\}\cap A=\emptyset$ \textbf{then} $R$
$\leftarrow$ $R$ $\cup$ (\qe$(u_1,A\cup\{u_1\})$
$\cap$ $\ldots$ $\cap$ \qe$(u_r,A\cup\{u_r\}))$ \\
4. \> \textbf{return}($R$)
\end{tabbing}
}\caption{The procedure \qe} \label{fig:esmmt}
\end{figure}

\begin{center}
\begin{table}
\caption{Evaluation of \qe$(a2,\{a2\})$ \label{tab:esmmt}}
\begin{tabular}{||r|l||} \hline\hline
Call & Result \\
\hline
\qe$(a2,\{a2\})$ & $I(a2)~\cup$ \qe$(b3,\{a2,b3\}) ~\cup$
(\qe$(b1,\{a2,b1\}) ~\cap$ \qe$(b2,\{a2,b2\}))$ \\
\qe$(b3,\{a2,b3\})$ & $I(b3)$ \\
\qe$(b1,\{a2,b1\})$ & $I(b1) ~\cup$ \qe$(c1,\{a2,b1,c1\}) ~\cup $
\qe$(c2,\{a2,b1,c2\})$ \\
\qe$(b2,\{a2,b2\})$ & $I(b2) ~\cup$ (\qe$(c2,\{a2,b2,c2\}) ~\cap$
\qe$(c3,\{a2,b2,c3\}))$ \\
\qe$(c1,\{a2,b1,c1\})$ & $I(c1)$ \\
\qe$(c2,\{a2,b1,c2\})$ & $I(c2)$ $\star$ \\
\qe$(c2,\{a2,b2,c2\})$ & $I(c2) ~\cup$ (\qe$(b1,\{a2,b2,c2,b1\}) ~\cap$
\qe$(b3,\{a2,b2,c2,b3\}))$ \\
\qe$(c3,\{a2,b2,c3\}))$ & $I(c3)$ \\
\qe$(b1,\{a2,b2,c2,b1\})$ & $I(b1) ~\cup$ \qe$(c1,\{a2,b2,c2,b1,c1\})$ $\star$
\\
\qe$(b3,\{a2,b2,c2,b3\}))$ & $I(b3)$ \\
\qe$(c1,\{a2,b2,c2,b1,c1\})$ & $I(c1)$ \\
\hline\hline
\end{tabular}
\end{table}
\end{center}

\vspace{-1.1cm}
The procedure \qe, presented in Figure~\ref{fig:esmmt}, computes $\ans(t,S)$ for
a given term $t$ (and an implicitly given source $S$) by applying in a
straightforward way Corollary~\ref{prop:iteli}.  To this end, \qe~must be
invoked as \qe$(t,\{t\}).$ The second input parameter of \qe~is the set of terms
on the \emph{path} from $t$ to the currently considered term $x.$ This set is
used to guarantee that $t$ is connected to all terms considered in the recursion
by a cycle-free simple path. \qe~accumulates in $R$ the result. The
correctness of \qe~can be established by just observing that, for all objects
$o\in\obj,$ $o$ is in the set $R$ returned by \qe$(t,\{t\})$ if and only if $o$
satisfies the two conditions expressed by Corollary~\ref{prop:iteli}.

As an example, let us consider the sequence of calls made by the procedure
\qe~in evaluating the query $a2$ in the example source of Figure
\ref{fig:extsource}, as shown in Table~\ref{tab:esmmt}.  The calls marked with a
$\star$ are those in which the test in line 3 gives a negative result.  Upon
evaluating \qe$(c2,\{a2,b1,c2\})$ the procedure realizes that the only incoming
hyperedge in $c2$ is $\langle \{b1,b3\},c2\rangle,$ whose tail $\{b1,b3\}$ has a
non-empty intersection with the current path $\{a2,b1,c2\};$ so the hyperedge is
ignored.  In this case, the cycle $(b1,c2,b1)$ is detected and properly handled.
Analogously, upon evaluating \qe$(b1,\{a2,b2,c2,b1\}),$ the cycle $(c2,b1,c2)$
is detected and properly handled. Also notice the difference between the calls
\qe$(c2,\{a2,b1,c2\})$ and \qe$(c2,\{a2,b2,c2\}).$ They both concern $c2,$ but
in the former case, $c2$ is encountered upon descending along the path
$(a2,b1,c2)$ whose next hyperedge is $\langle \{b1,b3\},c2\rangle;$ following
that hyperedge, would lead the computation back to the node $b1,$ which has
already been met, thus the result of the call is just $I(c2).$ In the latter
case, $c2$ is encountered upon descending along the path $(a2,b2,c2),$ thus the
hyperedge leading to $b1$ and $b3$ must be followed, since none of the terms in
its tail have been touched upon so far.

From a complexity point of view, \qe~visits all terms that lie on cycle-free
simple paths ending at the query term $t$ in the taxonomy B-graph $\ACCA.$ Now,
it is not difficult to see that these paths may be exponentially many in the
size of the taxonomy. As an illustration, let us consider the taxonomy whose
B-graph contains the following hyperedges:

\begin{center}
\begin{tabular}{lllll}
  $h_1:(\{u_1,v_1\},u_2)$ & $h_2:(\{u_2,v_2\},u_3)$ & $\ldots$ &
  $h_{n-1}:(\{u_{n-1},v_{n-1}\},u_{n})$ & $h_n:(\{u_n,v_n\},t)$ \\
  $g_1:(\{u_1,v_1\},v_2)$ & $g_2:(\{u_2,v_2\},v_3)$ & $\ldots$ &
  $g_{n-1}:(\{u_{n-1},v_{n-1}\},v_{n})$ &
\end{tabular}
\end{center}
Let us assume $t$ is the query term. It is easy to verify that there are
$2^{n-1}$ cycle-free simple paths connecting $u_1$ to $t,$ one for each sequence of
the form
\begin{center}
  $(u_1~f_1~x_2~f_2~\ldots~x_{n-1}~f_{n-1}~x_n~h_n~t)$
\end{center}
where $f_i$ can be either $h_i$ (in which case $x_{i+1}$ is $u_{i+1}$) or $g_i$
(in which case $x_{i+1}$ is $v_{i+1}$) for $1\leq i\leq n-1.$

On the other hand, for each query term, \qe~performs set-theoretic operations on
sets of objects, which initially are interpretations of terms. Thus, we conclude
that \qe~has polynomial time complexity w.r.t. the size of $\obj$.

\subsection{Networks of Information Sources}
\label{sec:Mediators}

In this Section we complete the definition of our model by introducing networks
of information sources.  In order to be a component of a networked information
system, a source is endowed with additional subsumption relations, called
articulations, which relate the source terminology to the terminologies of other
sources of the same kind.

\begin{definition}[Articulation] \label{def:Articulation}
  \emph{Given two terminologies $T$ and $U,$ an \emph{articulation} from $T$ to
    $U,$ is a pair $(q,t)$ where $q\in\EL_{U}$ is a conjunctive query and $t\in
    T.$} \ok
\end{definition}

An articulation is not syntactically different from a subsumption relationship,
except that its head may be a term of a different terminology than the one where
the terms making up its tail come from.

\begin{definition}[Articulated source] \label{def:ArticulatedSource}
  \emph{An {\em articulated source} $\ES$ over $k\geq 0$ disjoint terminologies
    $T_1,\ldots,T_k,$ is a 5-tuple $\ES=(T_\ES,\preceq_\ES,\obj,I_\ES,R_\ES),$
    where:
    \begin{itemize}
    \item $(T_\ES, \preceq_\ES,\obj,I_\ES)$ is a source;
    \item $R_\ES$ is a set of articulations $(q,t)$ where $t\in T_\ES,$ $q$ is a
      conjunctive query in $\EL_T$ and $T=\cup_{i=1}^k T_i.$ \ok
    \end{itemize}
  }
\end{definition}

Articulations are used to connect an articulated source to other articulated
sources, so creating a networked information system. An articulated source $\ES$
with an empty interpretation, \ie~$I_\ES(t)=\emptyset$ for all $t\in
T_\ES,$ is called a \emph{mediator} in the literature.

\begin{definition}[Network] \label{def:Network}
  \emph{A \emph{network of articulated sources,} or simply a \emph{network},
    $\N$ is a non-empty set of articulated sources $\N=\{\ES_1,\ldots,\ES_n\},$
    where each $\ES_i$ is articulated over the terminologies of some of the
    other sources in $\N$ and all terminologies $T_{\ES_1},\ldots,T_{\ES_n}$ of
    the sources in $\N$ are disjoint.}  \ok
\end{definition}

Notice that the domain of the interpretation of an articulated source is
independent from the source, thus the same for any articulated source. This is
not necessary for our model to work, just reflects a typical situation of
networked resources such as URLs. Relaxing this constrain would have no impact
on the results reported in the present study.

An intuitive way of interpreting a network is to view it as a single source
which is distributed along the nodes of a network, each node dealing with a
specific vocabulary. The global source can be logically constructed by removing
the barriers which separate local sources, as if (virtually) collecting all the
network information in a single repository.  The notion of \emph{network source}
captures this interpretation of a network.

\begin{definition}[Network source, network query] \label{def:NetworkSource}
  \emph{The \emph{network source} $S_\N$ of a network of articulated sources
    $\N=\{\ES_1,\ldots,\ES_n\},$ is the source
    $S_\N=(T_\N,\sqsubseteq_\N,\obj,I_\N),$ where:
    \begin{itemize}
    \item $T_\N = \bigcup_{i=1}^n T_{\ES_i};$
    \item $I_\N  =  \bigcup_{i=1}^n I_{\ES_i}$
    \item $\sqsubseteq_\N = \bigcup_{i=1}^n (\preceq_{\ES_i} \cup \;R_{\ES_i})$  
    \end{itemize}
    A \emph{network query} is a query over $T_\N.$ } \ok
\end{definition}

The source $S_\N$ emerges in a bottom-up manner from the articulations of the
sources, as postulated in~\cite{AbererEmergentSemantics04}.  Note that
Definition~\ref{def:ArticulatedSource} does not necessarily imply that
$\preceq_\ES,$ $R_\ES,$ and $I_\ES$ are stored in the articulated source $\ES.$
In fact, given a network of articulated sources $\N,$ in Section~\ref{sec:appr},
several architectures will be considered for storing $\s_\N$ and $I_\N.$ A
network query is a query in anyone of the languages of the sources making up the
network. As it will be evident, our query evaluation method only requires minor
modifications to be able to evaluate also queries in the language $\EL_{T_\N},$
that is queries that mix terms from different terminologies.

The answer to a network query $q,$ or \emph{network answer,} is given by
$\ans(q,S_\N).$

Figure~\ref{fig:net} presents the taxonomy of a network source $S_\N,$ where
$\N$ consists of 3 sources $\N=\{P_a,P_b,P_c\}.$ As it can be verified, this is
the same taxonomy as the one shown in Figure~\ref{fig:extsource}, except that
now some of its subsumption relationships are elements of articulations.

\begin{figure}[htbp]
  \centering
  \begin{picture}(6.5,3)
    \put(0.75,2.25){\makebox(0.5,0.5){$a1$}}
    \put(1.25,1.25){\makebox(0.5,0.5){$a2$}}
    \put(0.25,1.25){\makebox(0.5,0.5){$a3$}}
    \put(2.5,2.25){\makebox(1,0.5){$b1\wedge b2$}}
    \put(3.5,1.25){\makebox(1,0.5){$b1\wedge b3$}}
    \put(2.5,1.25){\makebox(0.5,0.5){$b3$}}
    \put(4,2.25){\makebox(0.5,0.5){$b1$}}
    \put(3.75,0.25){\makebox(0.5,0.5){$b2$}}
    \put(5.25,2.25){\makebox(0.5,0.5){$c1$}}
    \put(5.25,1.25){\makebox(0.5,0.5){$c2$}}
    \put(5.25,0.25){\makebox(1,0.5){$c2\wedge c3$}}

    \put(0.5,1.75){\vector(1,2){0.25}}
    \put(1.5,1.75){\vector(-1,2){0.25}}
    \put(2.5,2.25){\vector(-3,-2){0.75}}
    \put(2.5,1.5){\vector(-1,0){0.75}}
    \put(5.2,2.5){\vector(-1,0){0.75}}
    \put(5.25,1.65){\vector(-1,1){0.75}}
    \put(4.55,1.5){\vector(1,0){0.7}}
    \put(5.1,0.5){\vector(-1,0){0.85}}
    \put(1,2){\oval(2,2){}}
    \put(3.5,1.5){\oval(2.5,3){}}
    \put(5.75,1.5){\oval(1.5,3){}}
    \put(0,0.65){\makebox(0.5,0.35){$P_a$}}
    \put(1.9,0){\makebox(0.5,0.35){$P_b$}}
    \put(6.35,0){\makebox(0.5,0.35){$P_c$}}
  \end{picture}
  \caption{A network taxonomy}
  \label{fig:net}
\end{figure}

\section{Query Evaluation Principles}
\label{sec:ba}

Before delving into distributed and optimized architectures, we now put the
query evaluation problem in a software perspective, illustrating the basic
principles that will be followed in subsequent developments.

We begin by distinguishing between two basic approaches for carrying out query
evaluation:
\begin{itemize}
\item The Direct approach, in which the answer is computed in one stage.
\item The Rewriting (or two-stage) approach, in which query evaluation is
  performed in two stages: a \emph{re-write} of the query is computed in the
  first stage and evaluated in the second stage.
\end{itemize}
Each one of these approached will be illustrated in the rest of this Section in
a general way. The implementation details will be specified in the next
Section. In both cases, the processes performing the query evaluation task will
communicate in an asynchronous way, by exchanging messages through the
appropriate queues. In this way, no process is blocked waiting for some other
process to finish and the number of servers can be expanded at will. The former
fact favours efficiency, while the latter favours scalability.

\subsection{Direct query evaluation}
\label{sec:bad}

The main processes involved in direct query evaluation are:
\begin{itemize}
\item \query,
\item \procask, and
\item \proctell.
\end{itemize}

\subsubsection{\query}
\label{sec:query}

\query~has two main tasks: to handle the communication with applications and to
initiate query evaluation. Following the syntax for queries given in
Definition~\ref{def:QueryLang}, \query~receives in input queries $q$ of the
form:
\[
q = \bigvee C_i
\]
where each $C_i$ is a conjunctive query. As a first step, \query~reduces $q$ to
a term query $t$ by (a) generating a new term $t$ not in $T,$ and (b) inserting
a new hyperedge $(C_i, t)$ into the taxonomy B-graph for each conjunctive query
$C_i$ (see Proposition~\ref{prop:termqonly}). A new query id \emph{ID} for $t$
is subsequently obtained, and an \ask~message is sent (see below) for
evaluating $t,$ including \emph{ID}, $t,$ and the set of already visited terms,
that is just $t.$ Finally, \emph{ID} is returned, to allow the requesting
application to retrieve the query result as soon as it is available.

As an example, let us consider the network shown in Figure~\ref{fig:net}, whose
B-graph is given in Figure~\ref{fig:extsource} left, and the query $a2\wedge
a3$. When given as input to \query, a new term $t$ is generated and the
hyperedge $(\{a2,a3\},t)$ is added to the taxonomy B-graph.  The id of the new
query is also generated, let it be $q1.$ Then \query~sends the message
\ask($q1$,$t$,$\{t\}$) and returns $q1.$

\subsubsection{\procask}
\label{sec:ask}

An \ask~message represents the request of evaluating a term query and consists
of 3 fields:
\begin{itemize}
\item the id of the term query;
\item the term constituting the query;
\item the set of already visited terms (as requested by \qe).
\end{itemize}
The basic task of \procask~is to analyze an \ask~message in order to ascertain
whether there are hyperedges to consider for evaluating the given term query,
\ie~any hyperedge that passes the test on line 3 of \qe. If yes, \procask~breaks
down the query into sub-queries as established by \qe, and launches the
evaluation of these sub-queries by issuing the corresponding \ask~messages. If
not, it just returns the interpretation of the given term.

In our running example, upon processing the message ($q1$, $t$, $\{t\}$),
\procask~finds that the hyperedge $(\{a2,a3\},t)$ needs to be considered. This
hyperedge requires breaking the term query $q1$ into two sub-queries, one for
the term $a2$ and one for the term $a3.$ The intersection of the results of
these sub-queries will have to be computed in order to have the final
result. Now each sub-query needs a unique identifier; let us assume that $q2$ is
the identifier of the sub-query relative to term $a2,$ and $q3$ is that relative
to $a3.$ Then \procask~issues the following \ask~messages:
\begin{itemize}
\item ($q2,$ $a2,$ $\{t,a2\}$), and
\item ($q3,$ $a3,$ $\{t,a3\}$).
\end{itemize}
Notice that in each message the set of visited terms is expanded as established
by $\qe.$

In order to keep track of the evaluation of a query \emph{ID}, a \emph{query
  program} is associated to \emph{ID,} given by a set of \emph{sub-programs}
$\{\subp_1,\ldots,\subp_k\}$ where each sub-program $\subp_j$ is relative to a
hyperedge to be considered in the evaluation of \emph{ID,} and is given by a set
of \emph{calls.} A call represents a sub-query of \emph{ID,} and can be:
\begin{itemize}
\item \emph{open,} meaning that the sub-query is being evaluated, in which
  case the call is the sub-query id;
\item \emph{closed,} meaning the sub-query has been evaluated, in which case
  the call is the resulting set of objects.
\end{itemize}
A query program is \emph{closed} if all calls in it are closed. In the above
example, the program associated to $q1$ consists of just one sub-program (since
there is only one relevant hyperedge) given by $\{q2,q3\}.$

Upon processing the message ($q3,$ $a3,$ $\{t,a3\}$), \procask~finds that there
are no hyperedges incoming into term $a3.$ Thus the query can be evaluated
immediately, which \procask~does by issuing the \tell~message ($q3,$ $I(a3)$),
which just tells that the result of $q3$ is $I(a3).$

\begin{center}
  \begin{table}
  \caption{Messages generated in the direct evaluation of the query $a2$
    \label{tab:esmmtask}}
  \begin{tabular}{||r|l|l||} \hline\hline
    Incoming message & Generated messages & Q. Program \\
    \hline \ask$(1,a2,\{a2\})$ & \ask$(2,b3,\{a2,b3\}),$ \ask$(3,b1,\{a2,b1\}),$
    &  \\
    & \ask$(4,b2,\{a2,b2\}))$ & $\{\{2\},\{3,4\}\}$ \\
    \ask$(2,b3,\{a2,b3\})$ & \tell$(2,I(b3))$ & \\
    \ask$(3,b1,\{a2,b1\})$ & \ask$(5,c1,\{a2,b1,c1\}),$
    \ask$(6,c2,\{a2,b1,c2\})$ &
    $\{\{5\},\{6\}\}$ \\
    \ask$(4,b2,\{a2,b2\})$ & \ask$(7,c2,\{a2,b2,c2\}),$
    \ask$(8,c3,\{a2,b2,c3\}))$ &
    $\{\{7,8\}\}$ \\
    \ask$(5,c1,\{a2,b1,c1\})$ & \tell$(5,I(c1))$ & \\
    \ask$(6,c2,\{a2,b1,c2\})$ & \tell$(6,I(c2))$ & \\
    \ask$(7,c2,\{a2,b2,c2\})$ & \ask$(9,b1,\{a2,b2,c2,b1\}),$
    \ask$(10,b3,\{a2,b2,c2,b3\})$ & $\{\{9,10\}\}$ \\
    \ask$(8,c3,\{a2,b2,c3\}))$ & \tell$(8,I(c3))$ & \\
    \ask$(9,b1,\{a2,b2,c2,b1\})$ & \ask$(11,c1,\{a2,b2,c2,b1,c1\})$ & $\{\{11\}\}$ \\
    \ask$(10,b3,\{a2,b2,c2,b3\}))$ & \tell$(10,I(b3))$ & \\
    \ask$(11,c1,\{a2,b2,c2,b1,c1\})$ & \tell$(11,I(c1))$ & \\
    \tell$(11,I(c1))$ & \tell$(9, I(b1)\cup R(11))$ & \\
    \tell$(10,I(b3))$ & the query program of 7 becomes $\{\{9,R(10)\}\}$ & \\
    \tell$(9, I(b1)\cup R(11))$ & \tell$(7,I(c2)\cup(R(9)\cap R(10)))$ & \\
    \tell$(8,I(c3))$ & the query program of 4 becomes $\{\{7,R(8)\}\}$ & \\
    \tell$(7,I(c2)\cup(R(9)\cap R(10)))$ & \tell$(4,I(b2)\cup(R(7)\cap R(8)))$ &
    \\
    \tell$(6,I(c2))$ & the query program of 3 becomes $\{\{5\},\{R(6)\}\}$ & \\
    \tell$(5,I(c1))$ & \tell$(3,I(b1)\cup R(5) \cup R(6))$ & \\
    \tell$(4,I(b2)\cup(R(7)\cap R(8)))$ & the query program of 1 becomes
    $\{\{2\},\{3,R(4)\}\}$ & \\
    \tell$(3,I(b1)\cup R(5) \cup R(6))$ & the query program of 1 becomes
    $\{\{2\},\{R(3),R(4)\}\}$ & \\
    \tell$(2,I(b3))$ & \tell$(1,I(a2)\cup R(2)\cup(R(3)\cap R(4)))$ & \\
    \hline\hline
  \end{tabular}
\end{table}
\end{center}

\subsubsection{\proctell}
\label{sec:tell}

When a \tell~message (\qid, $R$) is processed, \qid~is a sub-query of some other
query \qidone, that is an open call in the query program associated to
\qidone. The basic task of \proctell~is to ascertain whether the result of
\qid~is the last one needed for computing the result of \qidone. If not,
\proctell~just records the result of \qid~by replacing \qid~by $R$ in the query
program of \qidone. In our example, upon processing the message ($q3,$ $I(a3)$),
\proctell~updates $q1$'s program which becomes $\{q2,I(a3)\}.$

If \qid~is the last open call, then the query program of \qidone~is closed, in
which case \proctell~computes the result of \qidone~and communicates it by
issuing a corresponding \tell~message. The result of a closed program
$\{\subp_1,\ldots,\subp_m\},$ where each sub-program $\subp_i$ is a collection
of object sets $\subp_i=\{R_1^i,\dots,R_{m_i}^i\},$ is the set of objects given
by:
\begin{equation}
  \label{eq:qp}
  \bigcup_{i=1}^m \;\bigcap_{j=1}^{m_i}\; R_j^i.
\end{equation}
Notice that the processing of a \tell~message may cause the issue of another
\tell~message, and so on, until eventually all the sub-query's programs of a query
are closed and the final answer is obtained.

The complete series of \ask~and \tell~messages produced during the evaluation of
the query $a2$ in the example source of Figure~\ref{fig:extsource} is given in
Table~\ref{tab:esmmtask}, whose columns show: the incoming message, the
generated messages (when no message is generated, the changes to the relevant
query program are reported), and the query program generated in
\ask~messages. Queries are identified by non-negative integers, while $R(n)$
stands for the result of query $n.$ This Table should be compared with
Table~\ref{tab:esmmt}, showing the sequence of \qe~calls for the same query
evaluation.

\subsection{Correctness and complexity}
\label{sec:cc}

The combined action of \procask~and \proctell~is equivalent to the behavior of
the procedure \qe. To see why, it suffices to consider the following facts:
\begin{enumerate}
\item An \ask~message is generated for each recursive call performed by \qe~and
  vice-versa, that is whenever \qe~would perform a recursive call, an
  \ask~message is generated.  Therefore, the number of \ask~messages is the same
  as the number of terms that can be found on all B-paths from $t.$
\item For each \ask~message, exactly one \tell~message results. This can be
  observed by considering that, for each processed \ask~message, there can be
  two cases:
  \begin{enumerate}
  \item there is no hyperedge to consider: in this case, no subsequent
    \ask~message is generated, and a \tell~message is generated;
  \item there is at least one hyperedge to consider: in this case a number of
    sub-queries is generated and evaluated by issuing the corresponding
    \ask~messages; each such message has a larger set of visited terms. Since
    the B-graph is finite, eventually each sub-query will lead to a term falling
    in the previous case (this is how \qe~terminates). When the program of all
    sub-queries of a given term query $t$ is closed, \proctell~issues a
    \tell~message on $t.$ This will propagate the closure upwards, until all
    open calls are closed.
  \end{enumerate}
\item Finally, a closed query program is interpreted by computing (see
  expression (\ref{eq:qp})) the same operation on the result of sub-queries as
  \qe~does on the results of its recursive calls.
\end{enumerate}
As a consequence of these facts, we have the correctness of the above described
network query evaluation procedure, and also its polynomial time complexity with
respect to the size of $\obj.$ Note that the total number of messages generated
is twice the number of terms visited by \qe, and the number of query programs is
no larger than that.

\subsection{Re-writing based query evaluation}
\label{sec:barw}

\begin{figure}
  \centering
  \begin{picture}(6,7.5)

    \put(1,7){\makebox(0.5,0.5){$\cup$}}
    \put(0,6){\makebox(0.5,0.5){$a2$}}
    \put(1,6){\makebox(0.5,0.5){$b3$}}
    \put(2,6){\makebox(0.5,0.5){$\cap$}}

    \put(0.5,6.5){\line(1,1){0.5}}
    \put(1.25,6.5){\line(0,1){0.5}}
    \put(2,6.5){\line(-1,1){0.5}}

    \put(1,5){\makebox(0.5,0.5){$\cup$}}
    \put(3,5){\makebox(0.5,0.5){$\cup$}}

    \put(1.5,5.5){\line(1,1){0.5}}
    \put(3,5.5){\line(-1,1){0.5}}

    \put(0,4){\makebox(0.5,0.5){$b1$}}
    \put(1,4){\makebox(0.5,0.5){$c1$}}
    \put(2,4){\makebox(0.5,0.5){$c2$}}
    \put(3,4){\makebox(0.5,0.5){$b2$}}
    \put(4,4){\makebox(0.5,0.5){$\cap$}}

    \put(0.5,4.5){\line(1,1){0.5}}
    \put(1.25,4.5){\line(0,1){0.5}}
    \put(2,4.5){\line(-1,1){0.5}}
    \put(3.25,4.5){\line(0,1){0.5}}
    \put(4,4.5){\line(-1,1){0.5}}

    \put(3,3){\makebox(0.5,0.5){$\cup$}}
    \put(5,3){\makebox(0.5,0.5){$c3$}}

    \put(3.5,3.5){\line(1,1){0.5}}
    \put(5,3.5){\line(-1,1){0.5}}

    \put(2,2){\makebox(0.5,0.5){$c2$}}
    \put(4,2){\makebox(0.5,0.5){$\cap$}}

    \put(2.5,2.5){\line(1,1){0.5}}
    \put(4,2.5){\line(-1,1){0.5}}

    \put(3,1){\makebox(0.5,0.5){$\cup$}}
    \put(5,1){\makebox(0.5,0.5){$b3$}}

    \put(3.5,1.5){\line(1,1){0.5}}
    \put(5,1.5){\line(-1,1){0.5}}

    \put(2,0){\makebox(0.5,0.5){$b1$}}
    \put(4,0){\makebox(0.5,0.5){$c1$}}

    \put(2.5,0.5){\line(1,1){0.5}}
    \put(4,0.5){\line(-1,1){0.5}}

  \end{picture}
  \caption{A re-write of the query shown in Table~\ref{tab:esmmt}}
  \label{fig:rew}
\end{figure}
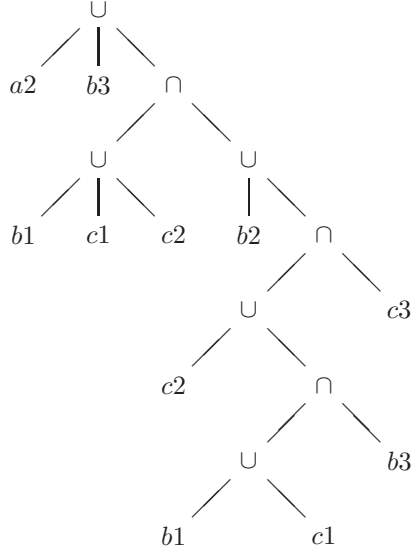

A query re-write represents in a symbolic way the computation of a query result
according to the procedure \qe. Specifically, a query re-write is a syntax tree
with 3 types of nodes: union nodes, intersection nodes and terms. The first two
types of nodes are non-terminal, whereas all terminal nodes are terms. To
evaluate a query re-write means to replace the terms by their interpretation and
then to execute the unions and the intersections as they appear in the syntax
tree, finally obtaining a set of objects. The re-write of the query whose
\qe~calls are presented in Table~\ref{tab:esmmt}, is given in
Figure~\ref{fig:rew}.

Query evaluation based on re-writing is a slight variation of the method
presented in Section~\ref{sec:ba}, in which \ask~messages and query programs are
exactly the same, while \tell~messages return linearizations of sub-trees of the
re-write; the last \tell~message returns the whole re-write in a linear form.

To exemplify, Table~\ref{tab:esmmttell} shows the \tell~messages produced in the
re-writing of the query $a2,$ whose direct evaluation is reported in
Table~\ref{tab:esmmtask}.

\begin{center}
  \begin{table}
  \caption{Messages generated in the re-writing of the query $a2$
    \label{tab:esmmttell}}
  \begin{tabular}{||r|l||} \hline\hline
    Incoming message & Generated messages \\  \hline
    \ask$(2,b3,\{a2,b3\})$ & \tell(2,``$b3$'') \\
    \ask$(5,c1,\{a2,b1,c1\})$ & \tell(5,``$c1$'') \\
    \ask$(6,c2,\{a2,b1,c2\})$ & \tell(6,``$c2$'') \\
    \ask$(8,c3,\{a2,b2,c3\}))$ & \tell(8,``$c3$'') \\
    \ask$(10,b3,\{a2,b2,c2,b3\}))$ & \tell(10,``$b3$'') \\
    \ask$(11,c1,\{a2,b2,c2,b1,c1\})$ & \tell(11,``$c1$'') \\
    \tell(11,``$c1$'') & \tell(9, ``$b1\cup R(11)$'') \\
    \tell(10,``$b3$'') & the query program of 7 becomes $\{\{9,R(10)\}\}$ \\
    \tell(9, ``$b1\cup R(11)$'') & \tell(7,``$c2\cup(R(9)\cap R(10))$'') \\
    \tell(8,``$c3$'') & the query program of 4 becomes $\{\{7,R(8)\}\}$ \\
    \tell(7,``$c2\cup(R(9)\cap R(10))$'') & \tell(4,``$b2\cup(R(7)\cap R(8))$'')
    \\
    \tell(6,``$c2$'') & the query program of 3 becomes $\{\{5\},\{R(6)\}\}$ \\
    \tell(5,``$c1$'') & \tell(3,``$b1\cup R(5) \cup R(6)$'') \\
    \tell(4,``$b2\cup(R(7)\cap R(8))$'') & the query program of 1 becomes
    $\{\{2\},\{3,R(4)\}\}$ \\
    \tell(3,``$b1\cup R(5) \cup R(6)$'') & the query program of 1 becomes
    $\{\{2\},\{R(3),R(4)\}\}$ \\
    \tell(2,``$b3$'') & \tell(1,``$a2\cup R(2)\cup(R(3)\cap R(4))$'') \\
    \hline\hline
  \end{tabular}
\end{table}
\end{center}

\section{Algorithms and Architectures for Network Query Evaluation}
\label{sec:appr}

We now consider distributed architectures for network query evaluation, based on
the approaches outlined in the previous Section. In order to identify all
significant architectures, it is important to consider how taxonomies and
interpretations are allocated on the network.  In this respect, there are four
possibilities:
\begin{itemize}
\item Both the network taxonomy and interpretation are centralized, that
  is allocated on one source, which is termed \emph{global server.}
\item The network taxonomy is centralized in the \emph{taxonomy server,} while
  each source holds its own interpretation.
\item Each source holds its taxonomy (including articulations), while the network
  interpretation is allocated to a single source, the \emph{interpretation
    server.}
\item Both the network taxonomy and interpretation are distributed to the
  sources, in a pure peer-to-peer model.
\end{itemize}
Considered in conjunction with the two evaluation approaches identified in the
previous Section, \ie~direct and re-writing, these possibilities give rise to 8
different architectures. In order to indicate any one of them, we will use
3-letter names, as follows: the first and second letters denote, respectively
the allocation of taxonomy and interpretation (C standing for centralized and D
for distributed), while the third letter indicates the type of evaluation (D for
direct, R for rewriting).  Thus, CDR denotes the approach in which the taxonomy
is centralized, the interpretations are distributed and the query is first
re-written and then evaluated. The rest of this study is devoted to rank these
methods with respect to their performance in terms of response time. In this
respect, some approaches stand out immediately as not particularly
promising. Namely,
\begin{itemize}
\item When there is a global server source, query re-writing (CCR) is clearly a
  looser with respect to the direct approach (CCD): if everything is in one
  place, there is no gain to be made in following a two-stage approach; as a
  consequence, the approach CCR will no longer be considered.
\item For the opposite reason, CDD is a clear looser with respect to CDR: if the
  taxonomy is centralized, in CDR the taxonomy server is contacted only once to
  re-write the query, while in CDD is invoked at every sub-query evaluation.
\item For the same reason, also DCD is a clear looser with respect to DCR: if
  the interpretation is centralized, it is more convenient to re-write the query
  and consult the interpretation server only once for the final evaluation,
  rather than invoking it repeatedly.
\end{itemize}
Before delving into the analysis of the remaining 5 methods, we present the
model of a source, which is common to all methods.

\subsection{Model of a source}
\label{sec:arch}
\begin{figure}[htbp]
  \centering
  \begin{picture}(12,6)
    \put(0,4){\framebox(2,1.5){Application}}
    \put(3.5,4){\framebox(2,1.5){
        \begin{minipage}{2cm}
          \begin{center}
            Query \\ Receiver
          \end{center}
        \end{minipage}
      }}
    \put(3.5,0.5){\framebox(2,1.5){
        \begin{minipage}{2cm}
          \begin{center}
            IRQ \\ Server
          \end{center}
        \end{minipage}
      }}
    \put(6.5,0.5){\framebox(2.5,3){
        \begin{minipage}{3cm}
          \begin{center}
            Source \\ Component
          \end{center}
        \end{minipage}
      }}
    \put(11,2){\oval(2,1)}
    \put(10,1.5){\makebox(2,1){Network}}


    \multiput(3.5,2.5)(1.5,0){2}{\line(0,1){1}}
    \multiput(3.5,2.75)(0,0.5){2}{\line(1,0){1.5}}
    \put(3.5,2.75){\makebox(1.5,0.5){IRQ}}
    \multiput(6.5,4.5)(1,0){2}{\line(0,1){1}}
    \multiput(6.5,4.75)(0,0.5){2}{\line(1,0){1}}
    \put(6.5,4.75){\makebox(1,0.5){AQ}}
    \multiput(8,4.5)(1,0){2}{\line(0,1){1}}
    \multiput(8,4.75)(0,0.5){2}{\line(1,0){1}}
    \put(8,4.75){\makebox(1,0.5){ORQ}}


    \put(2,4.5){\vector(1,0){1.5}}
    \put(3.5,5){\vector(-1,0){1.5}}
    \put(7,5.5){\vector(0,1){0.5}}
    \put(7,6){\line(-1,0){6}}
    \put(1,6){\vector(0,-1){0.5}}

    \put(8.5,5.5){\vector(0,1){0.5}}
    \put(8.5,6){\line(1,0){2.5}}
    \put(11,6){\vector(0,-1){3.5}}
    \put(7,3.5){\vector(0,1){1}}
    \put(8.5,3.5){\vector(0,1){1}}
    \put(5.5,1.25){\vector(1,0){1}}
    \put(4.5,4){\vector(0,-1){0.5}}
    \put(4,3.75){\vector(0,-1){0.25}}
    \put(4.5,2.5){\vector(0,-1){0.5}}
    \put(4,3.75){\line(-1,0){1}}
    \put(3,3.75){\line(0,-1){3.75}}
    \put(3,0){\line(1,0){8}}
    \put(11,1.5){\vector(0,-1){1.5}}

  \end{picture}
  \caption{Architecture of a Source}
  \label{fig:arch}
\end{figure}
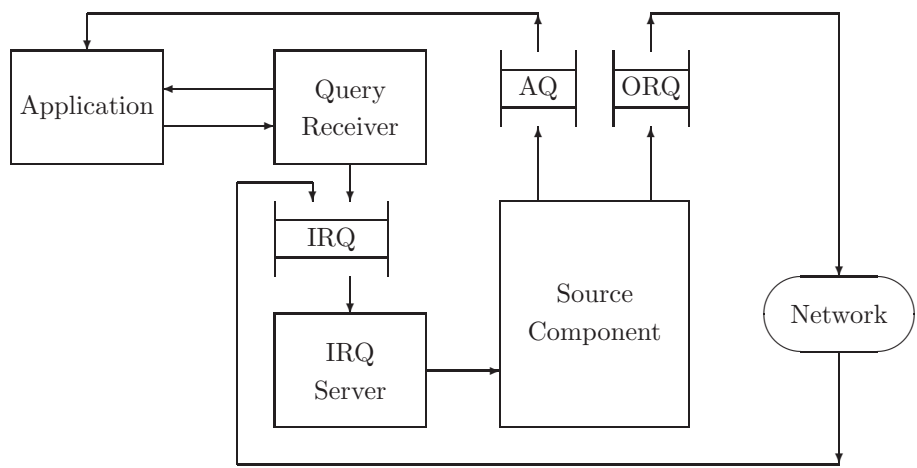

A source (Figure~\ref{fig:arch}) consists of three main architectural elements:
\begin{itemize}
\item \emph{Applications,} which formulate queries and wait to receive the
  corresponding answers.
\item \emph{Source Component,} which is a set of processes, exposed via an API,
  implementing query evaluation according to the principles outlined in the
  previous Section. As we will see, the methods exposed by a Source Component,
  as well as their semantics, may vary depending on the approach.
\item \emph{Communication Components,} consisting of the data structures and the
  processes which manage the interaction between the Source Component from one
  hand, and the network and the Applications from the other.  Inter-process
  communication is implemented by means of queues, as anticipated. The following
  queues are part of the architecture of every type of source: \emph{Input
    Request Queue} (IRQ), \emph{Output Request Queue} (ORQ) and \emph{Answer
    Queue} (AQ).  Query evaluation requests (whether from local applications or
  from other sources) are handled by the Query Receiver, which places them on
  the IRQ, from where the IRQ Server dequeues them for processing by the Source
  Component. Once a query is evaluated, the answer is placed on the AQ or on the
  ORQ, depending whether the request comes from a local application or another
  source, respectively. Messages posted on the ORQ of a source are directed to
  the IRQ of the receiving source. Due to the optimization techniques used for
  representing object identifiers and to the assumptions in query evaluation,
  messages are one-to-one with network packets, thus messages are the units of
  communication.
\end{itemize}

Every source uses a data structure, called \emph{Query Cache} (QC for short). QC
stores partial results and additionally works as a Cache, storing final results
for re-use. Two different time-outs are therefore used: the \emph{answer}
time-out ($t_a$), that is the amount of time until an answer is waited for; and
the \emph{cache} time-out ($t_c$), that is the amount of time an answer is
cached for re-use. At any time, an object in the QC is associated only with one
time-out, depending on its state.

A QC object corresponds to a query or a sub-query and has the following
attributes:
\begin{description}
\item[\id] the identifier of the object; we do not make any assumption on the
  structure of the identifier, except that the source where it has been created
  must be recoverable for it (for instance as an IP number);
\item[exp] the query expression; this may be the original query or a single
  term;
\item[state] the state of the object (see below);
\item[dependency] (\textbf{dep}, for short) the \id~of the oldest query $q$ into
  the QC having the same expression as the present query, if such a query
  exists; \texttt{null} otherwise; this attribute is kept for re-using the
  result of $q$ also for the present query, thus optimizing performance;
\item[answer] the answer of the query, if it has been computed; \texttt{null}
  otherwise;
\item[time-out] the expiration time of the object; after that point the object
  will eventually be deleted;
\item[QP] the query program associated to the object;
\item[n] the number of open calls in \textbf{QP};
\item[rewriting] a Boolean value set to \emph{true} if the first stage of the
  re-writing approach has been completed, to \emph{false} otherwise.
\end{description}

During its life inside the QC of a source, an object may be in one of the
following states (in what follows we will use ``query'' to mean the (sub)query
associated to the QC object):
\begin{description}
\item[\texttt{free}] the query is being evaluated, no answer has arrived for it
  so far, and no other query depends on it;
\item[\texttt{principal}] the query is being evaluated, no answer has arrived
  for it so far, and there exists at least one other query that depends on it;
\item[\texttt{dependent}] the query has been received but no evaluation for it
  has been launched, since there exists a \texttt{free}, \texttt{principal} or
  \texttt{declined} query with the same expression, which is being evaluated;
\item[\texttt{declined}] the query has expired before an answer for it was
  received, but it has not been deleted because other queries are dependent on
  it;
\item[\texttt{closed}] the answer for the query has been computed, and the query
  is kept in the Cache in order to re-use the answer, until it expires;
\item[\texttt{total}] the query is a term of an original query, it has to be
  evaluated, and the answer can be used to answer queries given by the same
  term.
\item[\texttt{partial}] the query is a term that has been encountered during the
  evaluation of a total term, thus its answer cannot be re-used to answer
  queries with the same expression.
\end{description}

In the re-writing approaches, two more states are defined:
\begin{description}
\item[\texttt{free-rw}] the query has been re-written and is being evaluated, no
  answer has arrived for it so far, and no other query depends on it;
\item[\texttt{principal-rw}] the query has been re-written and is being
  evaluated, no answer has arrived for it so far, and there exists at least one
  other query that depends on it.
\end{description}

\subsection{Query evaluation in the CCD Approach}
\label{sec:ccd-intro}

In the CCD approach, the network taxonomy and interpretation are
centralized (in Server Sources, see below), while the answer in computed in one
stage. This approach provides us with the basic concepts for describing the rest
of the architectures. Additionally, it provides us with the lowest bounds in our
performance evaluation of the different architectures.

In this approach, we can distinguish two different types of sources:
\emph{Client Source} and \emph{Server Source}, named after the fact that this is
indeed the classical client-server architecture.

\subsubsection{Client Source}
\label{sec:ccd-cl}
\label{sec:ccd-life-cs}
\label{sec:ccd-cqp}
\label{sec:ccd-cap}
\label{sec:ccd-ccm}

A Client Source (CS, for short) does not perform any one of the operations
involved in query evaluation. It receives queries from local applications and
simply sends them (via the ORQ) to a Server Source for evaluation; when the
corresponding answers arrive (in the IRQ), the CS makes them available to the
applications (in the AQ).

\begin{figure}[htbp]
  \centering
  \begin{picture}(12.5,7)
    \put(1.25,6.5){\oval(2.5,1)}
    \put(0,6){\makebox(2.5,1){Principal}}
    \put(1.25,4){\oval(2.5,1)}
    \put(0,3.5){\makebox(2.5,1){Free}}
    \put(8.75,6.5){\oval(2.5,1)}
    \put(7.5,6){\makebox(2.5,1){Closed}}
    \put(8.75,4){\oval(2.5,1)}
    \put(7.5,3.5){\makebox(2.5,1){Declined}}
    \put(8.75,0.5){\oval(2.5,1)}
    \put(7.5,0){\makebox(2.5,1){Dependent}}
    \put(1.25,1.75){\circle*{0.5}}
    \put(11,1.75){\circle{0.7}}
    \put(11,1.75){\circle*{0.5}}

    \put(2.5,6.5){\makebox(5,0.5){answer arrives}}
    \put(2.5,6.5){\vector(1,0){5}}
    \put(5,4.3){\makebox(2,1){
      \begin{minipage}{2cm}
        answer \\ time-out
      \end{minipage}
    }}
    \put(1.1,4.5){\makebox(2.5,1.5){
      \begin{minipage}{2cm}
        dependent \\ query \\ created
      \end{minipage}
    }}
  \put(1.25,4.5){\vector(0,1){1.5}}
  \put(1.1,2){\makebox(2.5,1.5){
      \begin{minipage}{2cm}
        no query \\ with same \\ expr. exists
      \end{minipage}
    }}
  \put(1.25,2){\vector(0,1){1.5}}
  \put(2.5,6.5){\vector(2,-1){5}}
  \put(2.5,4){\vector(1,1){2.5}}

  \put(7.6,4.5){\makebox(2,1.5){
      \begin{minipage}{2cm}
        answer \\ arrives
      \end{minipage}
    }}
  \put(8.75,4.5){\vector(0,1){1.5}}
  \put(1.25,1.5){\line(0,-1){1}}
  \put(1.25,0.5){\vector(1,0){6.25}}
  \put(1.5,0){\makebox(6.25,1){
      \begin{minipage}{6.25cm}
        free, principal or declined \\ query with same
      expression exists
      \end{minipage}
    }}
  \put(2.5,4){\vector(4,-1){8.2}}
  \put(5,2.8){\makebox(2,1){
      \begin{minipage}{2cm}
        answer \\ time-out
      \end{minipage}
    }}
  \put(9.8,0.9){\vector(2,1){1.05}}
  \put(10.1,0.2){\makebox(3,1){
      \begin{minipage}{3cm}
        ans. time-out or \\ answer arrives
      \end{minipage}
    }}
  \put(9.25,3.5){\vector(1,-1){1.5}}
  \put(9,2.5){\makebox(2,1){
      \begin{minipage}{3cm}
        no dependent \\ query exists
      \end{minipage}
    }}
  \put(1.5,1.75){\vector(1,0){9.15}}
  \put(3.3,1.7){\makebox(6,0.5){closed query with same expression exists}}
  \put(10,6.5){\line(1,0){1}}
  \put(11,6.5){\vector(0,-1){4.4}}
  \put(11.1,3.5){\makebox(2,1.5){
      \begin{minipage}{2cm}
        cache \\ time-out
      \end{minipage}
    }}

  \end{picture}
  \caption{Life-cycle of a QC object for the Client Source in the CCD Approach }
  \label{fig:ccd-life-cs}
\end{figure}
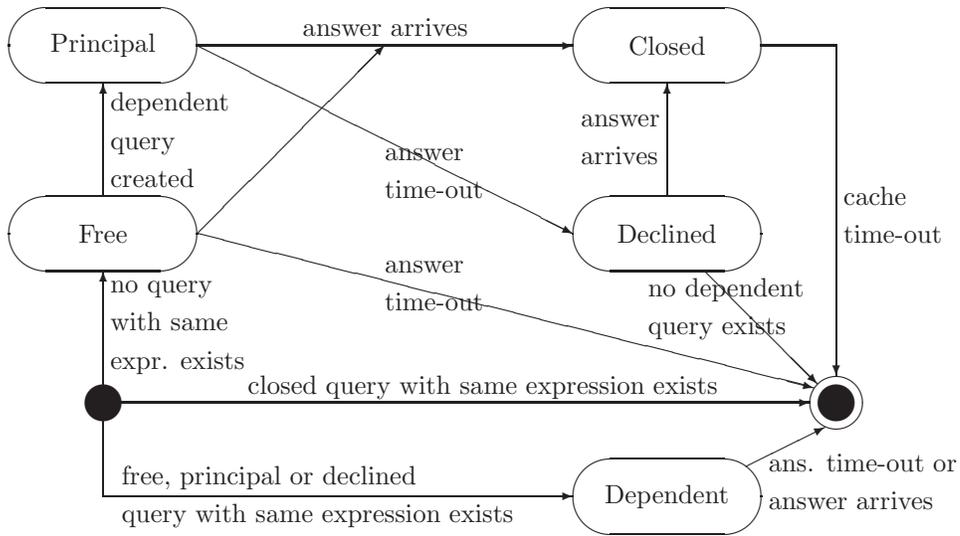

The state machine presented in Figure \ref{fig:ccd-life-cs} models the
life-cycle of a QC object in a Client Source. The QC of a CS only contains
objects corresponding to full queries, hence no \texttt{total} or
\texttt{partial} objects. We can distinguish three types of events: a new query
arrives; an answer arrives; and a time-out expires.  The arrival of a query
starts the life-cycle of a query. There may be three cases:
\begin{itemize}
\item No object exists having as expression the incoming query; in this case, a
  new object is created and its state is set to \texttt{free}.
\item A \texttt{closed} object $o$ having as expression the incoming query
  exists; in this case, the answer of $o$ is used for answering immediately the
  incoming query, and no new object is created.
\item A \texttt{free,} \texttt{principal} or \texttt{declined} object $o$ exists
  having as expression the incoming query; in this case, no evaluation needs to
  be launched for the incoming query, because a query with the same expression
  is being currently evaluated. Consequently, a new object is created, its state
  is set to \texttt{dependent} and its dependency is set to $o.$
\end{itemize}

A \texttt{free} object is deleted when its answer time-out expires; in this
case, the requesting application is notified that no answer for the query could
be computed. Otherwise, a \texttt{free} object becomes:
\begin{itemize}
\item \texttt{closed} when the corresponding answer arrives; or
\item \texttt{principal} if a dependency to it is created before the answer
  arrives.
\end{itemize}
A \texttt{principal} object, corresponding to a query not yet answered but
with other queries depending on it, becomes:
\begin{itemize}
\item \texttt{closed} when the corresponding answer arrives;
\item \texttt{declined} when the answer time-out expires. In this case, the
  requesting application is notified that no answer for the query could be
  computed but the object is not destroyed because the answer to the
  corresponding query is required to answer the queries of the dependent
  objects.
\end{itemize}
When the answer to a \texttt{declined} object arrives, the object becomes
\texttt{closed}, and the just arrived answer is used for answering all queries
dependent on it. The object is destroyed if all dependent objects expire (and
are destroyed) before the answer arrives; this is the only way such an object
dies; in other words, a \texttt{declined} object stays in the Cache until there
is a query dependent on it.

A \texttt{dependent} object can transition only into the final state, \ie\
be destroyed. This can happen in two different ways:
\begin{itemize}
\item if the answer time-out expires, the object is destroyed and a
  \texttt{null} answer is generated for it;
\item if the answer to the object on which it depends arrives, then that answer
  is used for it too.
\end{itemize}

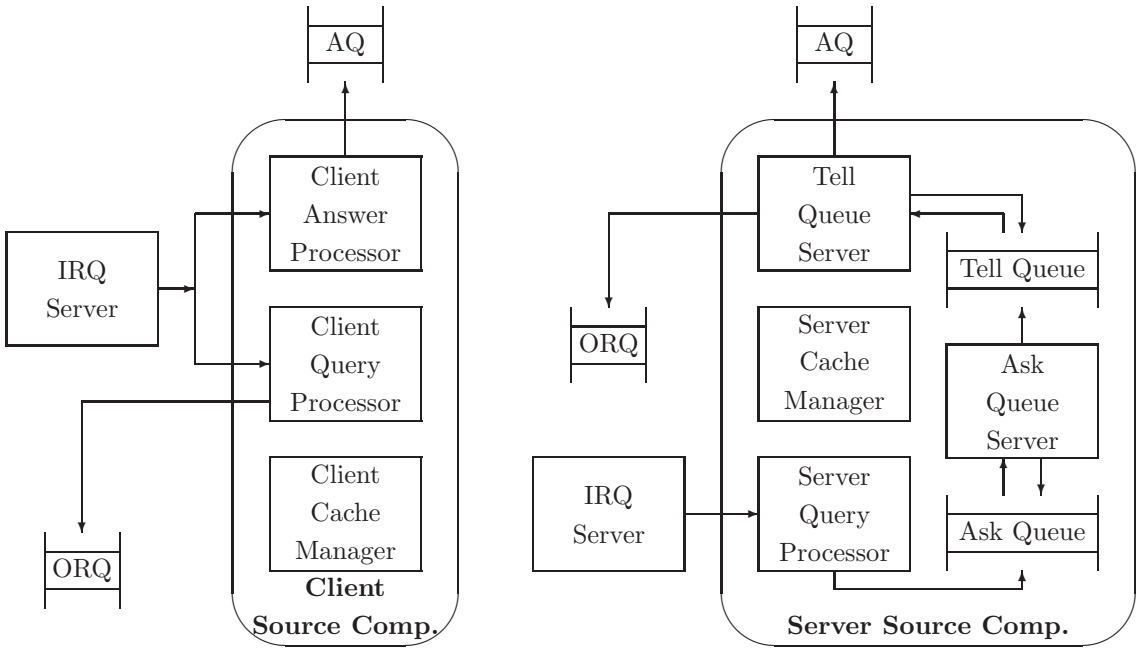
\begin{figure}[htbp]
  \centering
  \begin{picture}(15,8.5)


    \multiput(4,7.5)(1,0){2}{\line(0,1){1}}
    \multiput(4,7.75)(0,0.5){2}{\line(1,0){1}}
    \put(4,7.75){\makebox(1,0.5){AQ}}

    \put(4.5,3.5){\oval(3,7){}}
    \put(0,4){\framebox(2,1.5){
        \begin{minipage}{2cm}
          \begin{center}
            IRQ \\ Server
          \end{center}
        \end{minipage}
      }}
    \put(3.5,5){\framebox(2,1.5){
        \begin{minipage}{2cm}
          \begin{center}
            Client \\ Answer \\ Processor
          \end{center}
        \end{minipage}
      }}
    \put(3.5,3){\framebox(2,1.5){
        \begin{minipage}{2cm}
          \begin{center}
            Client \\ Query \\ Processor
          \end{center}
        \end{minipage}
      }}
    \put(3.5,1){\framebox(2,1.5){
        \begin{minipage}{2cm}
          \begin{center}
            Client \\ Cache \\ Manager
          \end{center}
        \end{minipage}
      }}
    \multiput(0.5,0.5)(1,0){2}{\line(0,1){1}}
    \multiput(0.5,0.75)(0,0.5){2}{\line(1,0){1}}
    \put(0.5,0.75){\makebox(1,0.5){ORQ}}

    \put(3,0){\makebox(3,1){
      \begin{minipage}{3cm}
        \begin{center}
          \textbf{Client \\ Source Comp.}
        \end{center}
      \end{minipage}
    }}


    \put(2,4.75){\vector(1,0){0.5}}
    \put(2.5,3.75){\line(0,1){2}}
    \put(2.5,3.75){\vector(1,0){1}}
    \put(2.5,5.75){\vector(1,0){1}}
    \put(4.5,6.5){\vector(0,1){1}}
    \put(3.5,3.25){\line(-1,0){2.5}}
    \put(1,3.25){\vector(0,-1){1.75}}


    \multiput(10.5,7.5)(1,0){2}{\line(0,1){1}}
    \multiput(10.5,7.75)(0,0.5){2}{\line(1,0){1}}
    \put(10.5,7.75){\makebox(1,0.5){AQ}}

    \put(12.25,3.5){\oval(5.5,7){}}
    \put(7,1){\framebox(2,1.5){
        \begin{minipage}{2cm}
          \begin{center}
            IRQ \\ Server
          \end{center}
        \end{minipage}
      }}
    \put(10,5){\framebox(2,1.5){
        \begin{minipage}{2cm}
          \begin{center}
            Tell \\ Queue \\ Server
          \end{center}
        \end{minipage}
      }}
    \put(10,3){\framebox(2,1.5){
        \begin{minipage}{2cm}
          \begin{center}
            Server \\ Cache \\ Manager
          \end{center}
        \end{minipage}
      }}
    \put(10,1){\framebox(2,1.5){
        \begin{minipage}{2cm}
          \begin{center}
            Server \\ Query \\ Processor
          \end{center}
        \end{minipage}
      }}
    \multiput(12.5,4.5)(2,0){2}{\line(0,1){1}}
    \multiput(12.5,4.75)(0,0.5){2}{\line(1,0){2}}
    \put(12.5,4.75){\makebox(2,0.5){Tell Queue}}
    \put(12.5,2.5){\framebox(2,1.5){
        \begin{minipage}{2cm}
          \begin{center}
            Ask \\ Queue \\ Server
          \end{center}
        \end{minipage}
      }}
    \multiput(12.5,1)(2,0){2}{\line(0,1){1}}
    \multiput(12.5,1.25)(0,0.5){2}{\line(1,0){2}}
    \put(12.5,1.25){\makebox(2,0.5){Ask Queue}}
    \multiput(7.5,3.5)(1,0){2}{\line(0,1){1}}
    \multiput(7.5,3.75)(0,0.5){2}{\line(1,0){1}}
    \put(7.5,3.75){\makebox(1,0.5){ORQ}}

    \put(9.5,0){\makebox(5.5,0.5){\textbf{Server Source Comp.}}}


    \put(9,1.75){\vector(1,0){1}}
    \put(11,1){\line(0,-1){0.25}}
    \put(11,0.75){\line(1,0){2.5}}
    \put(13.5,0.75){\vector(0,1){0.25}}
    \put(13.25,2){\vector(0,1){0.5}}
    \put(13.75,2.5){\vector(0,-1){0.5}}
    \put(13.5,4){\vector(0,1){0.5}}
    \put(13.25,5.75){\line(0,-1){0.25}}
    \put(13.25,5.75){\vector(-1,0){1.25}}
    \put(13.5,6){\line(-1,0){1.5}}
    \put(13.5,6){\vector(0,-1){0.5}}
    \put(10,5.75){\line(-1,0){2}}
    \put(8,5.75){\vector(0,-1){1.25}}
    \put(11,6.5){\vector(0,1){1}}

  \end{picture}
  \caption{Client and Server Source Components for the CCD approach}
  \label{fig:ccd-simple}
\end{figure}

Finally, a \texttt{closed} object stays in the QC until its cache time-out
expires. After that, it is destroyed (by the Cache Manager, see below).

The Source Component of a CS (Figure~\ref{fig:ccd-simple} left) includes three
main processes:
\begin{itemize}
\item Client Query Processor, invoked by the IRQ Server upon arrival of a new
  query on the IRQ. It handles this event as described above.
\item Client Answer Processor, invoked by the IRQ Server upon arrival of an
  answer message on the IRQ.
\item Client Cache Manager, which periodically inspects the QC in order to
  manage the expired objects.  Depending on the state of these objects,
  appropriate action is taken, as established by the query life-cycle.
\end{itemize}

\subsubsection{Server Source}
\label{sec:ccd-ssa}
\label{sec:ccd-life-ss}

A Server Source (SS, Figure~\ref{fig:ccd-simple} right) receives queries either
from the served Client Sources or from local applications (see
Section~\ref{sec:ba}); it evaluates these queries and sends the answers to the
appropriate requesters by placing them on either on the ORQ or on the AQ.

Four types of events can occur on a SS:
\begin{itemize}
\item A new query arrives.
\item A time-out expires.
\item An \ask~message arrives.
\item A \tell~message arrives.
\end{itemize}
The state machine modeling the life-cycle of a QC object in a SS is a simple
extension of that for a CS, and is not reported for brevity. The transitions
generated by the arrival of a new query are the same as those seen for a
CS. When the SS receives an \ask~message, a new object is generated, having as
state:
\begin{itemize}
\item \texttt{total}, if the involved term is a term in an original query; this
  can be ascertained by checking whether the set of the already visited terms
  has 2 elements;
\item \texttt{partial}, if the query is not total; this can be obviously
  ascertained by checking whether the set of already visited terms contains more
  than two terms.
\end{itemize}
If the answer time-out of a \texttt{total} or of a \texttt{partial} object
expires before the answer is received, a \tell~message with the special symbol
$\epsilon$ is generated; $\epsilon$ is interpreted as an empty answer not to be
cached. Notice that this preserves soundness of the query evaluation, while
giving up completeness.

Finally, when a non-$\epsilon$ \tell~message arrives for a \texttt{total}
object, the object becomes \texttt{closed} and the answer is stored to be used
to answer future queries, until it expires. On the other hand, when the
\tell~message relative to a \texttt{partial} object arrives, the object is
destroyed.

The Server Query Processor performs the operations of the \query~procedure
presented in the previous Section, reducing a full query to a term query, and
launching the execution of the latter via an \ask~message.

\ask~and \tell~messages are handled by the servers of the Ask and Tell Queues,
respectively, as shown in Figure~\ref{fig:ccd-simple} right. The Ask Queue
Server (AQS) is presented in Figure~\ref{fig:ccd-ask}.
\begin{figure}[htbp]
{\small
  \begin{tabbing}
    xxx \= xxx \= xxx \= xxx \= xxx \= \kill
     \procaqs \\
     1. \> \textbf{until} \askq~$\neq\emptyset$ \textbf{do} \\
     2. \> \> (\emph{ID}, $t$, $A$) $\leftarrow$ \textsc{Dequeue}(\askq); \\
     3. \> \> \textbf{if exists} $l \in \liq$ \textbf{such that}
     \textbf{exp}$[l]=t$ $\wedge$ \textbf{state}$[l]$=\clo~\textbf{then}\\
     4. \> \> \> \textsc{Enqueue}(\tellq,(\emph{ID}, \textbf{answer}$[l]$));\\
     5. \> \> \textbf{else if exists} $l \in \liq$ \textbf{such that}
     \textbf{\id}$[l]=ID$ \textbf{then} \\
     6. \> \> \> $n$ $\leftarrow$ $0;$ ~\qp,~$Q$ $\leftarrow$ $\emptyset$; \\
     7. \> \> \>\textbf{for each} hyperedge $h=\langle
     \{u_1,\ldots,u_r\},t\rangle$ \textbf{such that} $\{u_1,\ldots,u_r\}\cap
     A=\emptyset$ \textbf{do} \\
     8. \> \> \> \>$C$ $\leftarrow$ $\emptyset$; \\
     9. \> \> \> \>\textbf{for each} $u_i$ \textbf{do} \\
     10. \> \> \> \> \>$ID_i$ $\leftarrow$ (\emph{ID},\textsc{New-Num}); \\
     11. \> \> \> \> \>$C$ $\leftarrow$ $C$ $\cup$ \{$ID_i$\}; \\
     12. \> \> \> \> \>$n$ $\leftarrow$ $n+1$; \\
     13.\> \> \> \> \>\textsc{Enqueue}($Q,$ ($ID_i$, $u_i,$ $A\cup\{u_i\}))$; \\
     14.\> \> \> \>\qp~$\leftarrow$ \qp $\cup$ $\{C\}$; \\
     15.\> \> \> \textbf{if} $n>0$ \textbf{then}\\
     16.\> \> \> \>\textbf{n}$[l]$ $\leftarrow$ $n$; \textbf{QP}$[l]$
     $\leftarrow$ \qp;\\
     17.\> \> \> \>\textbf{until} $Q$ $\neq \emptyset$ \textbf{do} \\
     18.\> \> \> \> \>($I$, $u$, $B$) $\leftarrow$ \textsc{Dequeue}($Q$); \\
     19.\> \> \> \> \> $m$ $\leftarrow$ \textsc{new-query-cache};\\
     20.\> \> \> \> \>\textbf{\id}$[m]$ $\leftarrow$ $I$; \textbf{exp}$[m]$
     $\leftarrow$ $u$; \\
     21.\> \> \> \> \> \textbf{n}$[m]$ $\leftarrow$ $0$; \textbf{QP}$[m]$,
     \textbf{answer}$[m]$ $\leftarrow$ $\emptyset$; \textbf{dep}$[m]$
     $\leftarrow$ \emph{null}; \\
     22.\> \> \> \> \> \textbf{time-out}$[m]$ $\leftarrow$ \textsc{now}+$t_a$; \\
     23.\> \> \> \> \>\textbf{if} $|A|=2$ \textbf{then} \textbf{state}$[m]$
     $\leftarrow$ \tot~\textbf{else} \textbf{state}$[m]$ $\leftarrow$ \prt; \\
     24. \> \> \> \> \>\textsc{Enqueue}(\askq ,($I$, $u$, $B$)); \\
     25. \> \> \>\textbf{else} \textsc{Enqueue}(\tellq,(\emph{ID}, $I(t)$));
   \end{tabbing}
 }
 \caption{Ask Queue Server in CCD approach}
 \label{fig:ccd-ask}
\end{figure}
AQS implements the \procask~procedure of the previous Section. It dequeues the
first message from the \askq. Notice that at this point an object with
\textbf{Query-ID} attribute equal to \emph{ID} has already been created, either
by the Server Query Processor or by the AQS itself, but this object misses the
proper values of the \textbf{n} and \textbf{QP} attributes, which can be
computed only after analyzing the corresponding query. This analysis is carried
out now. AQS then checks (line 3) whether the Cache contains a \texttt{closed}
object $l$ whose query expression is the same term $t$ as the request being
processed. If such an object exists, its answer is used to answer the current
\ask~message, by creating a corresponding \tell~message and enqueuing it on the
\tellq~(line 4). If such an object does not exist, AQS checks (line 5) whether
the object corresponding to the query \emph{ID} still exists. If not, this
object has been destroyed because the answer time-out has expired, and in this
case AQS does nothing. If the object is found (in the variable $l$), AQS
examines the B-graph in order to find the hyperedges to be considered, \ie~those
which pass the test performed by \qe. If no hyperedge is found, then $n$ remains
0, the test on line 15 fails, and a \tell~message with $I(t)$ is put (line 25)
on the \tellq. Otherwise (lines 7 to 14), each hyperedge is processed by
generating a new sub-query for each term $u_i$ in its tail. The information
required to launch the execution of the sub-query (namely, its id, its term, and
the set of visited terms) is temporarily stored on a local queue $Q.$ The
sub-program corresponding to the hyperedge is accumulated on the variable $C,$
while $QP$ and $n$ store all generated sub-programs and the total number of open
calls, respectively. These values are assigned to the \textbf{QP} and \textbf{n}
attributes of $l$ thus completing the initialization of this object (line
16). Finally, AQS launches the evaluation of the generated sub-queries, in the
loop on lines 17-24. Until $Q$ is empty, it dequeues the information for
constructing an \ask~message for each sub-query, creating in the Cache a log
object $m$ representing the sub-query. $m$ is initialized in the lines 20-23 and
finally the corresponding sub-query is asked by putting an \ask~message on the
\askq.

\begin{figure}[hbtp]
{\small
  \begin{tabbing}
    Les \= xxx \= xxx \= xxx \= xxx \= xxx \= xxx \=\kill
    \proctqs\\
    1. \textbf{until} \tellq~$\neq\emptyset$ \textbf{do} \\
    2. \>(\emph{ID}, $R$) $\leftarrow$ \textsc{Dequeue}(\tellq);\\
    3. \> \textbf{if exists} $l \in \liq$ \textbf{such that}
    \textbf{Query-ID}$[l]=ID$ \textbf{then} \\
    4. \> \> \textbf{if state}$[l]$=\tot~\textbf{and} $R\neq\epsilon$
    \textbf{then do} \textbf{state}$[l]$ $\leftarrow$ \clo; \textbf{answer}$[l]$
    $\leftarrow$ $R$; \textbf{time-out}$[l]$ $\leftarrow$ \textsc{now}+$t_c$; \\
    5. \> \> \textbf{else} \textsc{delete}(\liq,~$l$);\\
    6. \> \> \textbf{if} $R=\epsilon$ \textbf{then} $R\leftarrow\emptyset$; \\
    7. \> \> \textbf{let} $l_1\in\liq$ \textbf{such that} \emph{ID} occurs in
    \textbf{QP}$[l_1]$; \\
    8. \> \> \qpone $\leftarrow$ \textsc{Close}(\textbf{QP}$[l_1]$,
    \textbf{\id}$[l_1]$, $R$); \\
    9. \> \> \textbf{if} \textbf{n}$[l_1]>1$ \textbf{then do}  \textbf{n}$[l_1]$
    $\leftarrow$ \textbf{n}$[l_1]-1$; \textbf{QP}$[l_1]$ $\leftarrow$ \qpone; \\
    10. \> \> \textbf{else do} \\
    11. \> \> \> $S$ $\leftarrow$ \textsc{Compute-answer}(\qpone);\\
    12. \> \> \> \textbf{if} \textbf{exp}$[l_1]$ $\in$ $T_{\mathit{self}}$ \textbf{then}
    \textsc{Enqueue}(\tellq, (\textbf{\id}$[l_1]$, $S\cup I(\mathbf{exp}[l_1])$);\\
    13. \> \> \> \textbf{else do} \\
    14. \> \> \> \> \textbf{if} \textbf{state}$[l_1]\neq\dec$ \textbf{then} \\
    15. \> \> \> \> \> \textbf{if} \textsc{extract-pid}(\emph{ID})=\emph{self}
    \textbf{then} \textsc{Enqueue}(\aq,(\textbf{Query-ID}$[l_1]$, $S$)); \\
    16. \> \> \> \> \> \textbf{else}
    \textsc{Enqueue}(\orq,(\textbf{Query-ID}$[l_1]$, $S$)); \\
    17. \> \> \> \> \textbf{if state}$[l_1]=\pri$ or \textbf{if
      state}$[l_1]=\dec$ \textbf{then} \\
    \> \> \> \> \> \textbf{for each} $l'$ $\in$ \liq~\textbf{such
      that} \textbf{dep}$[l']=ID$ \textbf{do} \\
    18. \> \> \> \> \> \> \textbf{if}
    \textsc{extract-pid}(\textbf{Query-ID}$[l']$)=\emph{self} \textbf{then}
    \textsc{Enqueue}(\aq,(\textbf{Query-ID}$[l']$, $S$)); \\
    19. \> \> \> \> \> \> \textbf{else}
    \textsc{Enqueue}(\orq,(\textbf{Query-ID}$[l']$, $S$)); \\
    20. \> \> \> \> \> \> \textsc{delete}(\liq, $l'$) \\
    21. \> \> \> \> \textbf{state}$[l_1]$ $\leftarrow$ \clo; \textbf{answer}$[l_1]$
    $\leftarrow$ $S$; \textbf{time-out}$[l_1]$ $\leftarrow$ \textsc{now}+$t_c$.
  \end{tabbing}
  }
  \caption{Tell Queue Server in CCD approach}
  \label{fig:ccd-tell}
\end{figure}

The \proctqs~(TQS) procedure is presented in Figure~\ref{fig:ccd-tell}.  First,
a \tell~message is dequeued. At this stage, an object $l$ with \textbf{Query-ID}
equal to \emph{ID} has been created in the QC (by AQS) and \emph{ID} is also an
open call in the query program of some other object $l_1.$ TQS manages both $l$
and $l_1.$ In particular, TQS has to check whether the \tell~message being
processed completes $l_1$'s evaluation, in which case TQS must issue the
relative \tell~message. Moreover, if $l_1$ is associated to an original query,
TQS must properly handle the answer and the state of $l_1.$

TQS first checks whether $l$ still exists in the QC. If it does not, then
nothing is done. Otherwise, TQS checks whether the current answer can be
re-used, which means that $l$'s state is \texttt{total} and $R$ is not the
special symbol $\epsilon.$ If this is indeed the case, $l$ becomes
\texttt{closed} and is properly updated (line 4); if not, $l$ is destroyed (line
5). On line 6, the meaning of $\epsilon$ as the empty answer in installed in
$R.$ Then, TQS retrieves $l_1$ (line 7) and uses \textsc{Close} to modify the
query program~\qp~in it, by closing the open call \qid: this means to replace
\qid~by $R,$ obtaining a new query program \qpone~(line 8). Then, the number of
open calls of $l_1$ is tested: if there are still open calls, $l_1$ evaluation
is not complete, thus its \textbf{n} and \textbf{QP} attributes are updated and
the procedure terminates. If the test on line 9 fails, \ie\ $n=1,$ then the
evaluation of the query corresponding to $l_1$ is completed, therefore the
result is computed in $S$ and it is tested whether the query term is in the
terminology of the source (line 12). If yes, the \emph{ID} is a sub-query,
therefore the obtained result is stored on a \tell~message which is enqueued on
the Tell Queue. If the query term of $l_1$ is not in the terminology of the
source, then it is a dummy term hence \qidone~is the id of an original query $q$
whose evaluation has been just completed.  In this case TQS must also manage the
object $l_1.$ If the state of $l_1$ is \texttt{declined} then the answer
time-out of this object has expired, thus the just computed answer is no longer
usable for its query. In all the other cases, the answer must be returned,
either locally (if \emph{PID} is the id of the present source, line 15) or
remotely (line 16). If the state of $l_1$ is \texttt{principal} or
\texttt{declined}, then other queries are depending on $l_1.$ Each object $l'$
relative to one dependent query is identified (line 17) and deleted (line 20)
after the corresponding answer is output either in the AQ (if local, line 18) or
in the ORQ (if remote, line 19). Finally, $l_1$ is updated (line 21) to be
subsequently re-used, until it expires.

\subsection{Query evaluation in the DDD Approach}
\label{sec:ddd-intro}

From an architectural point of view, DDD is the pure P2P approach, in which all
sources are of the same kind, each one storing its own taxonomy and associated
interpretation.

A DDD Source receives queries on its terminology from local applications, or
\ask~messages from other sources which, due to articulations, need to evaluate
some sub-query on the local terminology. The Source Component carries out the
evaluation of these queries or \ask~messages by relying on its taxonomy and
interpretation. Whenever it requires an answer to a sub-query outside its
terminology, it asks the appropriate source, from which it receives the
corresponding answer in a \tell~message.

The life-cycle of a query in a DDD Source is identical to that in a CCD Server
Source.

A DDD Source Component includes 4 main processes: Query Processor, Ask
Processor, Tell Processor, and Cache Manager.

\paragraph{Query Processor}
\label{sec:ddd-qp}

Query Processor (QP for short) is invoked by the Input Request Queue Server
whenever a Query message is dequeued having as fields (\emph{ID, q}). It
performs the following operations:
\begin{itemize}
\item If no query exists in the Cache with expression equal to $q,$ then QP
  behaves like a CCD Server: reduces $q$ to a term query $t,$ creates a new
  \texttt{free} query in the Cache and puts an \ask~message into the Input
  Request Queue, in order to launch the evaluation of $t.$
\item If there exists a \texttt{closed} query in the Cache with expression equal
  to $q,$ then QP uses the answer to that query to create an answer message for
  the input query, which it then puts in the Answer Queue.
\item If there exists a query $q'$ with expression equal to the input query and
  its state is neither \texttt{dependent} nor \texttt{closed}, then QP behaves
  like a CCD Client: creates a new \texttt{dependent} (from $q'$) query into the
  Cache and if $q'$ is \texttt{free}, QP sets it to \texttt{principal}.
\end{itemize}

\paragraph{Ask Processor}
\label{sec:ddd-ask}

The DDD Ask Processor (AP) behaves like the CCD Ask Queue Server, except that
the \ask~messages regarding terms belonging to other sources' terminologies are
inserted into the Output Request Queue (hence into the network) rather than into
the Ask Queue. As already pointed out, the \ask~messages posted on the ORQ will end
up in the IRQ of the receiving source, from where the IRQ Server dequeues them,
and uses their content to invoke the local AP.

\paragraph{Tell Processor}
\label{sec:ddd-tell}

The DDD Tell Processor (TP) behaves like the CCD Tell Queue Server, except that the
messages regarding the evaluation of sub-queries requested by other sources are
inserted into the Output Request Queue (hence into the network). Through the
previously described path, these messages will be processed by the TP of the
receiving Source.

\subsection{Query evaluation in the CDR Approach}
\label{sec:cdr-intro}

In a CDR architecture, the taxonomy is centralized but every source has its own
interpretation concerning the local terminology. This is a hybrid P2P approach,
which applies, for instance, when a central authority controls the vocabulary
used by a community of speakers for indexing their objects.  As a consequence,
there exist two types of sources: \emph{Source} and \emph{Server Taxonomy
  Source}.

\subsubsection{CDR Source}
\label{sec:cdr-csa}

The CDR Source component consists of 5 main processes: Query Processor, Query
Program Processor, Answer Processor, Local Interpretation Processor, and Cache
Manager.

The Query Processor receives queries from local applications. It sends each
query to a server taxonomy source to obtain the re-write of the query. When the
re-written query arrives, it is handled by the Query Program Processor, which
evaluates it by retrieving the interpretation for local terms, while
asking the appropriate sources for the interpretation of the foreign
terms. These requests are handled by the Local Interpretation Server of the
involved Sources. The answers to these requests are handled by the Answer
Processor, which eventually computes the query answer and makes it available to
the requesting application.

The life-cycle of a QC object in a CDR Source extends that of a CCD Client
Source, in order to manage the event of the arrival of a query re-write,
performed by the Query Program Processor as described below.

\paragraph{Query Program Processor}
\label{sec:cdr-qpp}

When a message containing a query re-write (\emph{ID}, \emph{rw}) arrives, the
Query Program Processor (QPP) checks whether the Cache contains an object $l$
whose \textbf{Query-ID} attribute value is \emph{ID.} If not, the object has
expired and has been destroyed, so nothing is done. If yes, QPP performs the
following operations:
\begin{itemize}
\item if the state of $l$ is \texttt{free} or \texttt{principal}, it changes it
  to \texttt{free-rw} or \texttt{principal-rw}, respectively; if the state of
  $l$ is \texttt{declined}, the object remains in the same state. In all these
  cases, the \qp~attribute is updated with the re-write, thereby caching the
  re-write for re-use.
\item it launches the evaluation of \emph{rw} as follows:
\begin{itemize}
\item it groups the terms in \emph{rw} by the source they belong;
\item it retrieves the interpretations of the local terms and inserts
  them into an answer message;
\item it requests  interpretations of the foreign terms by sending a
  message to the appropriate sources.
\end{itemize}
The grouping is very important, because it avoids requesting the same source
more than once.
\end{itemize}

\paragraph{Query Processor}
\label{sec:cdr-qp}

The Query Processor behaves like the Query Processor in the CCD Client Source,
except that when the Cache contains an object $l$ with the same expression as
the received one and whose state is not \texttt{closed} or \texttt{dependent},
QP creates a new \texttt{dependent} query object that depends on $l$ and if the
state of $l$ is \texttt{free} or \texttt{free-rw}, QP changes it to
\texttt{principal} or \texttt{principal-rw,} respectively.

\paragraph{Answer Processor}
\label{sec:cdr-ap}

When an answer message (\emph{ID,} \emph{A}) arrives containing the previously
requested interpretation(s) of foreign term(s), the Answer Processor (AP) checks
whether the QC contains an object $l$ whose \textbf{Query-ID} attribute value is
\emph{ID.} If not, the object has expired and has been destroyed; in this case,
AP does nothing.  If $l$ is found in the QC, then the sets of objects contained
in the answer message are used to replace the corresponding terms in the \qp~of
$l;$ if every term in \qp~has been replaced, then the answer to the query is
computed; moreover, the time-out attribute of $l$ is updated to cache time-out
and the query object is \texttt{closed}.  In addition:
\begin{itemize}
\item if the query is \texttt{free-rw}, the answer is output by putting a
  message in the Answer Queue;
\item if the query is \texttt{principal-rw}, the answer is output for the
  present and for all the depending queries;
\item if the query is \texttt{declined}, an answer is output only for all
  depending queries.
\end{itemize}

\paragraph{Local Interpretation Server}
\label{sec:cdr-lis}

The Local Interpretation Server is invoked when a message requesting the
interpretation of a set of terms belonging to the local terminology, arrives. It
retrieves the interpretation of every requested term, puts the result in
an answer message, and places the message into the Output Request Queue.

\subsubsection{Server Taxonomy Source}
\label{sec:cdr-stsa}

A Server Taxonomy Source (STS) carries out three basic tasks: the re-write of
all queries that it receives; the evaluation of the queries that it receives
from local applications; and the evaluation of the terms in its terminology
requested by remote sources. The re-writing stage is carried out locally as
described in Section~\ref{sec:barw}, which means that all \ask~and \tell~messages
are local.  Instead, the second stage of query evaluation implies the exchange
of \ask~and \tell~messages with the sources holding the foreign terms, and this
time \ask~and \tell~messages are as described in Section~\ref{sec:bad}.

The life-cycle of a QC object in a Server Taxonomy Source extends that of the
CCD Server Source with the management of the event concerning the arrival of a
re-write request for a query. During this stage, \texttt{partial} or
\texttt{total} QC objects are generated and evolved accordingly.  When the final
\tell~message arrives for a \texttt{total} query object, the object becomes
\texttt{free-rw} and its \qp~value is saved to be re-used for re-writing queries
with the same expression. On the other hand, when the final \tell~message to a
\texttt{partial} query object arrives, the object is destroyed.

\paragraph{Server Taxonomy Query Processor}
\label{sec:cdr-stqp}

The Server Taxonomy Query Processor (STQP) is invoked when a message requesting
to re-write a query $q$ arrives from a local application or from a remote
source.  STQP initiates the query re-write by placing an appropriate
\ask~message onto the Ask Queue, similarly to the Query Processor in the DDD
approach, with the following difference: when it creates a new
\texttt{dependent} object in the Cache, if the state of the referred object is
\texttt{free} or \texttt{free-rw}, STQP changes it to \texttt{principal} or
\texttt{principal-rw}, respectively.

 \paragraph{Ask Queue Server and Tell Queue Server}
 \label{sec:cdr-ask}

 These Servers behave in the same way as in the CCD architecture, except that
 \tell~messages contain linearizations of a re-write, as explained in
 Section~\ref{sec:barw}.

\paragraph{Answer Processor}
\label{sec:cdr-tell}

The Answer Processor (AP) in a STS performs the job of the Answer Processor and
the Query Program Processor in a CDR Source. Thus, an AP receives either a query
re-write to evaluate, or the interpretation of a set of foreign terms.

\subsection{Query evaluation in the DCR Approach}
\label{sec:dcr-intro}

This approach is dual to CDR. It includes two types of sources: \emph{Source} and
\emph{Server Interpretation Source}.  The life-cycle of a query in both types of
sources is identical to that of the CDR Server Taxonomy Source.

\paragraph{DCR Source}
\label{sec:dcr-cl}

A DCR Source does not have any interpretation, it only has the taxonomy
of the terms in its own terminology and articulations. When it receives queries
from local applications, it cooperates with other sources in order to carry out
the re-writing stage. In particular, it sends \ask~messages for the re-writing
of foreign terms, and receives \ask~messages for the re-writing of its own
terms. \tell~messages flow correspondingly. When the re-writing stage is
completed, the source sends the obtained query re-write to a Server
Interpretation Source for evaluation.

\paragraph{Server Interpretation Source}
\label{sec:dcr-stsa}

A Server Interpretation Source (SIS) has its own taxonomy and the whole network
interpretation. It carries out the re-writing of queries in cooperation with the
other sources, and in addition evaluates query re-writes.

\subsection{Query evaluation in the DDR Approach}
\label{sec:ddr-intro}

Similarly to the DDD approach, there is only one type of Source in DDR. A DDR
Source receives queries from local applications, and answer, \ask, \tell, and
interpretation request messages from other sources.  The Source carries out both
the first and the second stage of re-write based query evaluation method in
cooperation with the other sources.

\section{Performance Evaluation}
\label{sec:perfev}

In order to evaluate and compare the algorithms described in the previous
Section from a performance point of view, a simulation experiment has been run
for each of the 5 methods, using the same underlying network and under the same
query flow. The results of this experiment are summarized in
Table~\ref{tab:perftab}, which also indicates how long it took to obtain a
stable average response time in each case. The rest of this Section is devoted
to a description of the way these results have been obtained, and a discussion
on their meaning.

\begin{table}
  \centering
  \begin{tabular}{||c|c|c|c||}
    \hline\hline
    \emph{Method} &
    \begin{minipage}{2cm}
      \begin{center}
        \emph{Time to \\ stabilize \\ (minutes)}
      \end{center}
    \end{minipage}
    &
    \begin{minipage}{3cm}
      \begin{center}
        \emph{Avg response time \\ per retrieved  object \\ (milliseconds)}
      \end{center}
    \end{minipage}
    &
    \begin{minipage}{4cm}
      \begin{center}
        \emph{St. dev. response time\\ per retrieved  object \\ (milliseconds)}
      \end{center}
    \end{minipage}
    \\
    \hline
    CCD & 515 & 21.512 & 1.472 \\
    CDR & 445 & 25.301 & 0.139 \\
    DCR & 660 & 32.799 & 0.313 \\
    DDR & 650 & 34.336 & 0.251 \\
    DDD & 660 & 42.423 & 1.132 \\
    \hline\hline
  \end{tabular}
  \caption{Performance evaluation of the 5 methods}
  \label{tab:perftab}
\end{table}

\subsection{The Network Model}

The models of Source used for the simulation are exactly the same as those
presented in the previous Section. Thus, all types of Sources are structured as
illustrated in Figure~\ref{fig:arch}, and differ from one another in the Source
Component, which consists of the specific processes that have been illustrated
in the previous Section, for each approach.

\begin{figure}[htbp]
  \centering
  \leavevmode
  \centerline{\epsfig{figure=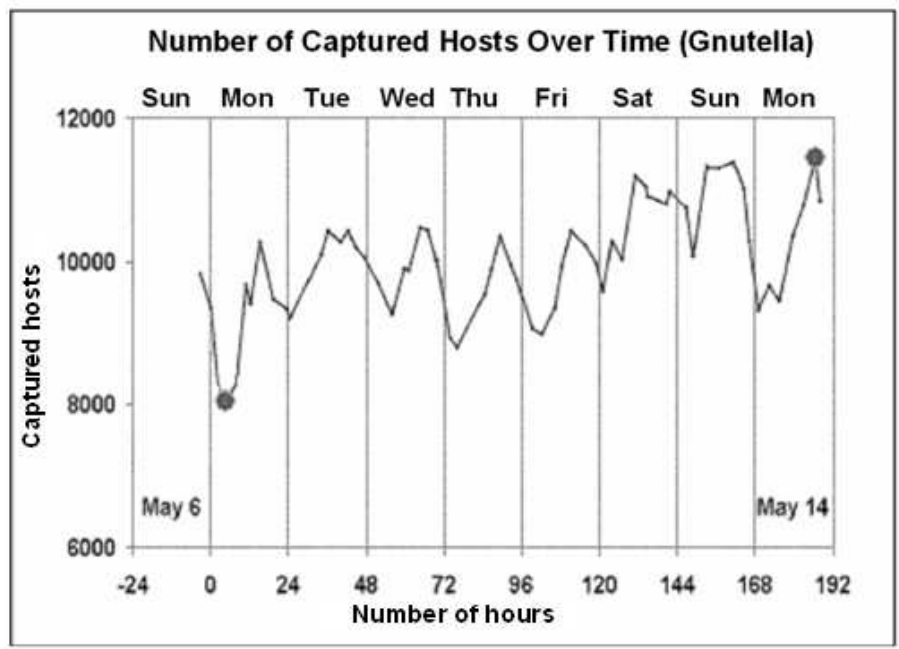, scale=0.9}}
  \caption{Statistical distribution of captured Gnutella peers}
  \label{fig:statdis}
\end{figure}

The data for configuring the underlying communication network, as well as other
important parameters, have been taken from statistical investigations on the
Internet, carried out at the University of Washington in the context of studies
on peer-to-peer file sharing~\cite{veronica2,veronica3} (see
Figure~\ref{fig:statdis}). This information has been used to estimate the delay
of operations performed by the TCP protocol and also the statistical
distribution of queries over time. The total delay is structured as follows:
\begin{itemize}
\item Queue Delay, depending on the degree of congestion of the network and the
  size of the involved Queue;
\item Processing Delay, given by the time required for: creating messages and
  decomposing/recomposing them into/from packets, and local query execution. It
  has been assumed that it takes $10^{-6}$ seconds to process one packet, and
  that a disk access takes 6 milliseconds (ms).
\item Packet Transmission Delay, proportional to the size of the packet and the
  bandwidth;
\item Propagation Delay, this is the network latency.
\end{itemize}
In order to measure the network latency we have used the estimations done on the
Gnutella network, given by the RTT (round-trip time) of a 40-byte TCP packet
exchanged between a peer and the measurement host. In particular, we have used
the following distribution:
\begin{itemize}
\item 20\% of peers have a latency smaller than 70 ms;
\item 20\% of peers have a latency higher than 280 ms;
\item the remaining 60\% of peers have latency uniformly distributed in between
  70 and 280 ms.
\end{itemize}
Also to estimate the network bandwidth we have relied on the measurements done
on Gnutella, according to which 78\% of the users are connected on a large
bandwidth (Cable, DSL, T1 or T3); according to the same estimation, about 30\%
of the users have a connection bandwidth higher than 3Mbps.

The size of the network was chosen to be 11400 sources, which is the maximum
number of simultaneously connected peers in Gnutella over a continuous period of
192 hours.

The number of servers utilized for CCD, CDR and DCR approaches is 57, that is
0.5\% of the total number of sources. Clearly, all the servers store the same
network taxonomy or interpretation (or both).

In order to obtain the taxonomy, interpretation, and articulations of each
source, the following parameters have been used, all with a uniform distribution:
\begin{itemize}
\item the size of the terminology is between 1 and 500 terms;
\item the size of the interpretation of any term is between 1 and 100 objects;
\item every source is articulated with  1 to 4 other sources;
\item the size of a local taxonomy is 25\% the size of the corresponding terminology;
\item the total number of articulations is 6\% of the size of the network terminology.
\end{itemize}

We have assumed that objects are URLs. This is typical in P2P
networks. Following~\cite{veronica6}, the average size of a URL is 63.4
bytes. The size of the internal representation of a term is assumed to be the
minimum amount of space required to uniquely identify an object within a set
(the terminology where the term belongs) on a network of sources, each identified
by an IP number. Finally, the values of the time-out are as follows: answer
time-out (the time a source waits for an answer to arrive) is 60 seconds, while
the cache time-out (the time a source keeps an answer in the Cache for re-use) is
600 seconds.

Most of these parameters are configurable.

\subsection{The simulation experiments}

In each experiment, the following variables have been measured:
\begin{itemize}
\item the time required to evaluate a query;
\item the number of evaluated queries, required to compute the average response time;
\item the size of the query result;
\item the number of packets transmitted, including the packets exchanged for the
  ping-pong protocol, which has been used for inter-source communication;
\item the number of visited sources.
\end{itemize}

For every approach, we have run a simulation experiment for the amount of time
required to obtain a stable value for the response time. Each query is
characterized by the following attributes:
\begin{itemize}
\item an integer number that identifies the query;
\item the id of the source that formulates the query, randomly chosen in the
  interval [1, 11400];
\item a set of terms randomly chosen from the terminology of the selected
  source, to be understood as a conjunction;
\item the time at which the query is issued.
\end{itemize}
The query distribution is obtained from the statistical distribution of
connected peers reported in Figure~\ref{fig:statdis}. In particular, the number
of queries per unit of time is directly proportional to the number of connected
peers. Since the same random generators have been used in all experiments for
generating the query parameters, the same query distribution is used in each
experiment.

In order to reduce the size of the required storage, we have gathered statistics
for periods of 5 minutes. The average response time was divided by the size of
the result; it was considered to be stable when the difference between the
values for 3 consecutive periods of 5 minutes, equivalent to 15 minutes of
simulated time, was less that $10^{-2}$ milliseconds. The goodness of this
choice is confirmed by the low value of the standard deviation for all evaluated
methods.

\begin{figure}[htbp]
  \centering
  \leavevmode
  \centerline{\epsfig{figure=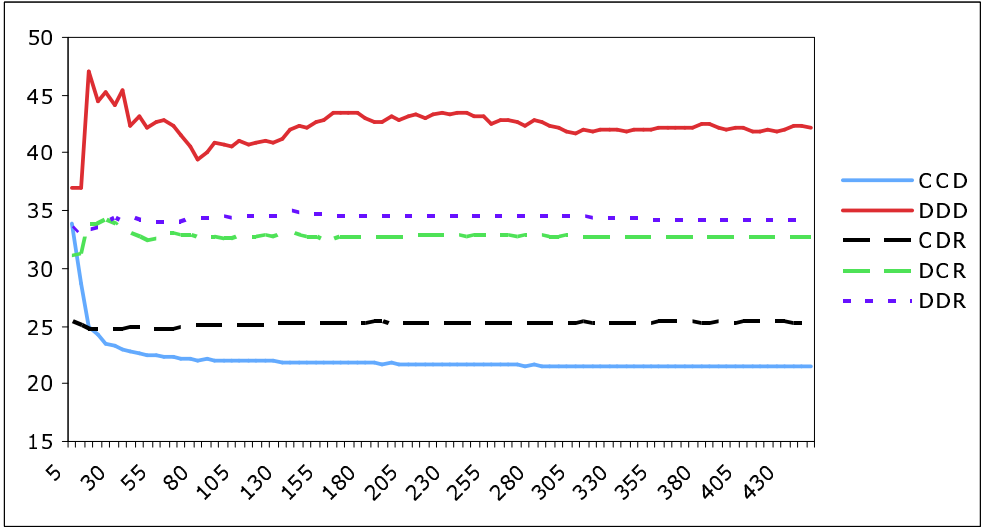, scale=1}}
  \caption{Average response time (milliseconds) per simulated
    time (minutes)}
  \label{fig:resptime}
\end{figure}

\subsection{Results and discussion}

Figure~\ref{fig:resptime} details the distribution of the average response time
in time for each method. As this Figure and Table~\ref{tab:perftab} show, the
fastest method is not surprisingly CCD, that is direct evaluation when both
taxonomy and interpretation are centralized. Perhaps more surprisingly, the next
best method is CDR, for reasons that will be analyzed below. As expected, the
worst method is DDD, which does not allow any type of optimization, and at each
step of the execution algorithm sends all collected terms and objects. The
performance of the DCR and DDR methods tend to be the same, the former being
slightly better because it can count on a centralized interpretation.

Needless to say, the actual values of the measured variables are determined also
by the parameters chosen to configure the simulator, detailed above. What really
matters are the relative values between the 5 compared methods, or in other
words, the ranking of the methods produced by the experiment.

\begin{figure}[htbp]
  \centering
  \leavevmode
  \centerline{\epsfig{figure=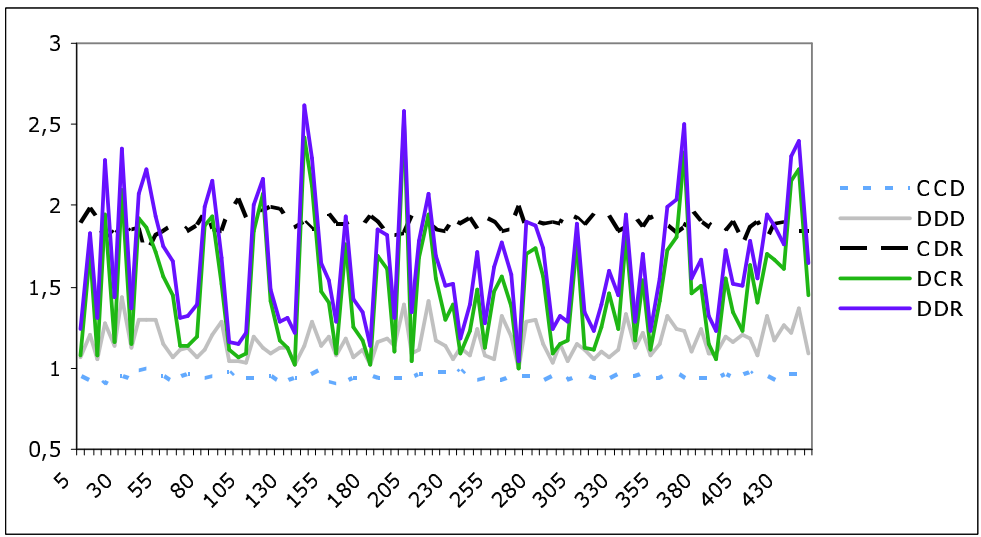, scale=1}}
  \caption{Average number of visited sources per simulated time
    (minutes)}
  \label{fig:visits}
\end{figure}

In order to verify the correctness of the results, for every pair of measured
variables the correlation coefficient has been computed. Correlation has been
confirmed between:
\begin{itemize}
\item number of generated sub-queries and the number of exchanged messages,
\item number of exchanged messages and number of visited sources.
\end{itemize}
In contrast, the number of exchanged messages and the size of the answer are
independent.

The average number of sources visited for evaluating one query is reported in
Figure~\ref{fig:visits}. The queries that either addressed the terminologies of
sources with no articulations or could be answered using the Cache, were
evaluated locally, and therefore considered to visit no sources. This explains
why the numbers in Figure~\ref{fig:visits} are so low, CCD being almost
everywhere below 1. As it can be seen, re-writing methods are those involving a
higher number of visits, with DDR having the highest for obvious reasons. Not
surprisingly, CCD is the method requiring the least number of visits of
all. However, the graphic suggests another clustering: the methods in which the
taxonomy is distributed have similar, very irregular curves, whilst those in
which the taxonomy is centralized have similar, much more regular curves. This
explains why CDR is the second best method after CCD: the centralization of the
taxonomy allows to re-write the query by contacting at most one source, the
taxonomy server; if instead the taxonomy is distributed, several sources may
need to be contacted in the re-writing stage, with the possibility that the same
source be contacted more than once, if articulations require so.  The
distribution of the interpretation may require to contact several sources for
the second stage, but the fact that this second stage is optimized implies that
every involved source is contacted exactly once, and this makes this step
affordable, so that on average 2 sources need to be visited, with a very small
standard deviation (in fact, for CDR the average number of visited source is
1.887, and the standard deviation is 0.053).

\begin{figure}[htbp]
  \centering
  \leavevmode
  \centerline{\epsfig{figure=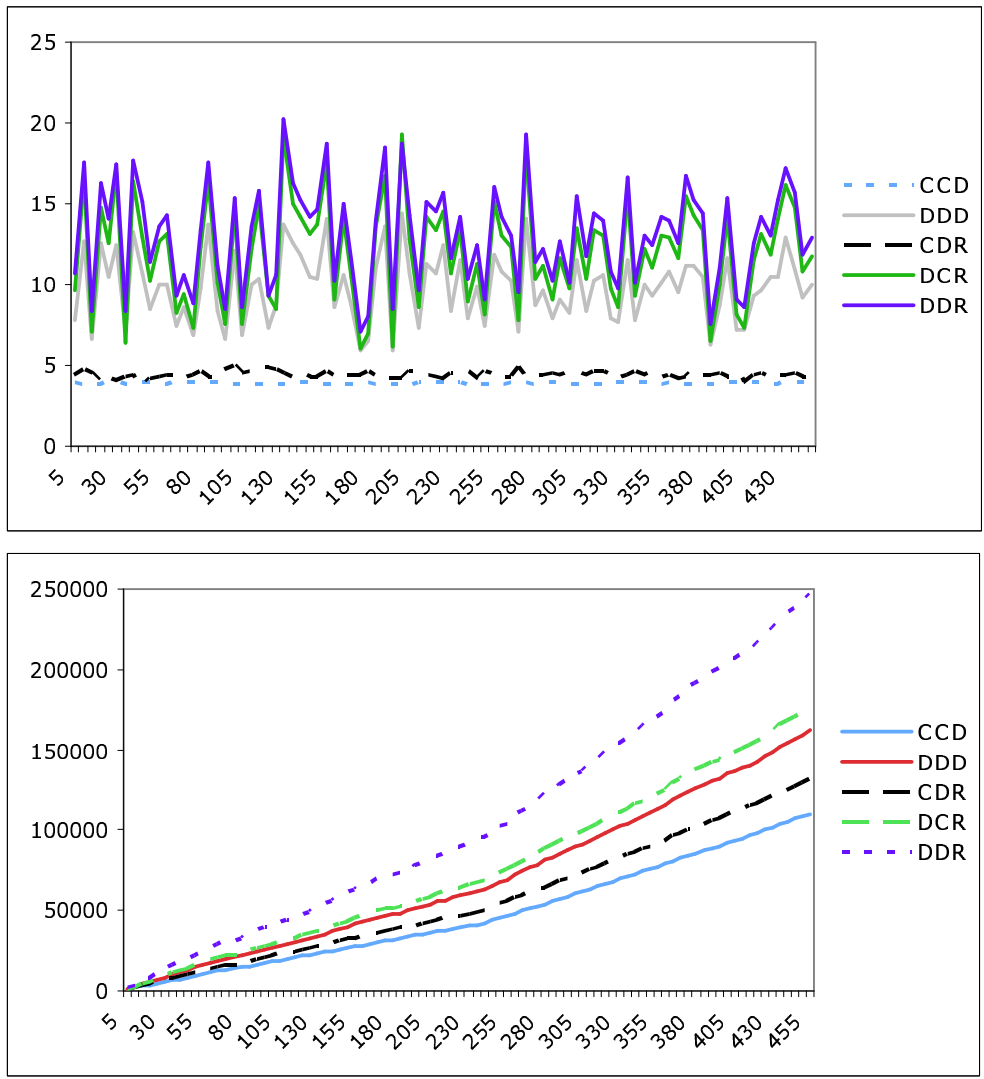, scale=1}}
  \caption{Average (top) and total (bottom) number of exchanged
    messages per simulated time (minutes)}
  \label{fig:msgs}
\end{figure}

This is confirmed by the number of exchanged messages
(Figure~\ref{fig:msgs}). As expected the methods in which the taxonomy is
distributed are those which require more message exchanges, with DDR being the
worse due to the fact that also interpretations are distributed and the
re-writing approach is followed. Also in this case CDR, although very similar to
DDR, does much better, up to being not significantly different from the best
method CCD.

Thus we can conclude that CDR is superior to all other distributed approaches
because the optimization taking place between the 2 stages of the re-writing
compensates the fact that more than one source needs to be contacted due to the
distribution of the interpretations. In other words, distributing the taxonomy
affects the performance of the method in a significant way, whilst optimization
can compensate the distribution of the interpretations.

\section{Related work}
\label{sec:rw}
In this Section we relate our work with the literature on peer-to-peer systems
and data integration, with emphasis on the former. Some parts of the work
reported in this paper have been already published.
Namely,~\cite{TzitzikasMeghiniER03} presents a first model of a network of
articulated sources, while \cite{TzitzikasMeghiniCoopIS03} studies query
evaluation on taxonomies including only term-to-term subsumption relationships.
Finally,~\cite{ArticulatedODBASE04} introduced \qe (without proving its
soundness and completeness), and gives hardness results for language extensions.

\paragraph{Description of our work in relation with P2P systems}
A peer-to-peer (P2P) system is a distributed system in which participants (the
peers) rely on one another for service, rather than solely relying on dedicated
and often centralized servers. The most popular P2P systems have focused on
specific application domains like music file sharing
\cite{Napster01,Gnutella,Kazaa}) or on providing file-system-like capabilities
\cite{Bolosky00}. In most of the cases, these systems do not provide
semantic-based retrieval services as the name of an object (e.g. the title of a
music file) is the only means for describing the object contents.

In our work, we make a distinction between the logical model of a network
(presented in Section~\ref{sec:found}) and the architecture for implementing
this model, which may be considered as a physical model of the
network. Typically, this distinction is not made in the literature on P2P
systems, thus we can compare only our pure P2P architecture, \ie~DDD, with the
P2P literature.

In the DDD approach, in order to evaluate a query $q$ posed to a peer $\ES$, the
incoming query (which is always expressed over its own taxonomy) is propagated
only to those peers to which $\ES$ has an articulation, and which can therefore
contribute to the answer of the query (the latter is determined by the taxonomy
and the articulations of $\ES$).  Specifically, it is not the original query to
be propagated, but a set of term sub-queries, each one belonging to the
terminology of the recipient peer. Note that in DDD there is not any form of
centralized index (like in Napster \cite{Napster01}), nor any flooding of
queries (like in Gnutella \cite{Gnutella}), nor any form of partitioned global
index (like in Chord \cite{Stoica01} and CAN \cite{CAN01}). Instead, we have a
query propagation mechanism that is query- and articulation-dependent (note that
Semantic Overlay Networks \cite{SONS2002} is a very simplistic approach to
this). Moreover note that the peers of our DDD model are quite autonomous in the
sense that they do not have to share or publish their stored objects, taxonomies
or mappings with the rest of the peers (neither to one central server, nor to
the on-line peers). To participate in the network, a peer just has to answer the
incoming queries by using its local base, and to propagate queries to those
peers that according to its ``knowledge'' (i.e.  taxonomy + articulations) may
contribute to the evaluation of the query. However both of the above tasks are
optional and at the ``will'' of the peer.

From a data modeling point of view several approaches for P2P systems have been
proposed recently, including relational-based approaches \cite{Bernstein02},
XML-based approaches \cite{Halevy03b} and RDF-based \cite{Edutella02b}.  In this
paper we consider the fully heterogeneous conceptual model approach (where each
peer can have its own schema), with the only restriction that each conceptual
model is represented as a taxonomy.  A taxonomy can range from a simple
tree-structured hierarchy of terms, to the concept lattice derived by Formal
Concept Analysis \cite{Ganter99}, or to the concept lattice of a Description
Logics theory. This taxonomy-based conceptual modeling approach has three main
advantages (for more see \cite{TzitzikasMeghiniER03}): (a) it is very easy to
create the conceptual model of a source, (b) the integration of information from
multiple sources can be done easily, and (c) automatic articulation using
data-driven methods (like the one presented in \cite{TzitzikasMeghiniCIA03}) are
possible.

Recently, there have been several works on P2P systems endowed with logic-based
models of the peers' information bases and of the mappings relating them (called
P2P mappings). These works can be classified in 2 broad categories: (1) those
assuming propositional or Horn clauses as representation language or as a
computational framework, and (2) those based on more powerful formalisms.  With
respect to the former category (\eg, see~\cite{rousset1,rousset2,adjman06}), our
work makes an important contribution, by providing a much simpler algorithm for
performing query answering than those based on resolution.  Indeed, we do rely
on the theory of propositional Horn clauses, but only for proving the
correctness of our algorithm.  For implementing query evaluation, we devise an
algorithm that avoids the (unnecessary) algorithmic complications that plague
the methods based on resolution. As an example, after appropriate
transformations our framework can be seen as a special case of that
in~\cite{rousset2}. Then, query evaluation can be performed by first computing
the prime implicates of the negation of each term in the query, using the
resolution-based algorithms presented in~\cite{rousset2}. As the complexity of
this problem is exponential w.r.t the size of the taxonomy and polynomial
w.r.t. the size of $\obj,$ there is no computational gain in using this
approach. Instead, there is an algorithmic loss, since the method is much more
complicated than ours.

As for the second category above, works in this area have focused on providing
highly expressive knowledge representation languages in order to capture at once
the widest range of applications. Notably, \cite{CalvanesePODS04} proposes a
model allowing, among other things, for existential quantification both in the
bodies and in the heads of the mapping rules.  Inevitably, such languages pose
computational problems: deciding membership of a tuple in the answer of a query
is undecidable in the framework proposed by~\cite{CalvanesePODS04}, while
disjunction in the rules' heads makes the same problem coNP-hard already for
datalog with unary predicate (\ie~terms).  These problems are circumvented in
both approaches by changing the semantics of a P2P network, in particular by
adopting an epistemic reading of mappings. As a result, the inferential relation
of the resulting logic is weakened up to the point of making the above mentioned
decision problem solvable in polynomial time.  In contrast, our approach aims at
reaching the same goal (efficient support for P2P information access), but from
a different standpoint: in particular, we achieve efficiency of the network
query evaluation by limiting the expressiveness of the language for representing
mappings (articulations, in our terminology), while retaining a classical
Tarskian semantics of these mappings (seen as logical formulae). In other words,
we aim at a smaller class of applications, but for this we offer a framework
resting on the classical logical foundations. The complementarity of the two
approaches is therefore evident.  Since we have also
shown~\cite{ArticulatedODBASE04},~\cite{TzitzikasMeghiniCoopIS03} that classical
semantics leads to intractability as soon as the expressiveness of the mapping
language is increased, we can say that we have covered a large part of our side
of the trade-off.

In \cite{CalvaneseDBISP2P-2003}, a query answering algorithm for simple P2P
systems is presented where each peer $\ES$ is associated with a local database,
an (exported) peer schema, and a set of local mapping rules from the schema of
the local database to the peer schema. P2P mapping rules are of the form $cq_1
\leadsto cq_2$, where $cq_1, cq_2$ are conjunctive queries of the same arity $n
\geq 1$ (possibly involving existential variables), expressed over the union of
the schemata of the peers, and over the schema of a single peer,
respectively\footnote{Note that P2P mapping rules of this kind can accommodate
  both GAV and LAV-style mappings, and are referred in the literature as GLAV
  mappings.}.  Note that this representation framework partially subsumes our
network source framework, since in our case $cq_1, cq_2$ are of arity 1, $cq_1$
is a conjunctive query of the form $u_1(x) \wedge\ldots\wedge u_r(x)$ over the
terminology of a single peer\footnote{Recall that this restriction can be easily
  relaxed.}  and $q_2$ is a single atom query $t(x)$ over the terminology of the
peer that the mapping (articulation) belongs to.  However, simple P2P systems
cannot express the local to a peer $\ES$ taxonomy $\preceq_\ES$ of our
framework.  Query answering in simple P2P systems according to the first-order
logic (FOL) semantics is in general undecidable. Therefore, the authors adopt a
new semantics based on epistemic logic in order to get decidability for query
answering.  Notably, the FOL semantics and epistemic logic semantics for our
framework coincide.  In particular, in \cite{CalvaneseDBISP2P-2003}, a
centralized bottom-up algorithm is presented which essentially constructs a
finite database $RDB$ which constitutes a ``representative'' of all the epistemic
models of the P2P system.  The answers to a conjunctive query $q$ are the
answers of $q$ w.r.t. $RDB$.  However, though this algorithm has polynomial time
complexity, it is centralized and it suffers from the drawbacks of bottom-up
computation that does not take into account the structure of the query.

The work in \cite{CalvaneseDBISP2P-2003} is extended in
\cite{CalvanesePODS04}, where a more general framework for P2P systems
is considered, which fully subsumes our framework and whose semantics
is based on epistemic logic.  In particular, in
\cite{CalvanesePODS04}, a peer is also associated with a set of
(function-free) FOL formulas over the schema of the peer. A top-down
distributed query answering algorithm is presented which is based on
synchronous messaging. Essentially, the algorithm returns to the peer
where the original query is posed, a datalog program by transferring
the full extensions of the relevant to the query, peer source
predicates along the paths of peers involved in query processing. The
returned datalog program is used for providing the answers to the
query. Obviously, our algorithm has computational advantages w.r.t.
the algorithm in \cite{CalvanesePODS04}, since during query evaluation
only the full or partial answer to a term (sub)query is transferred to
the peer that posed the (sub)query, and not the full extensions of all
terms involved in its evaluation.

The framework in \cite{SerafChid00}, extends our framework by
considering (i) $n$-ary (instead of unary) predicates (\ie~P2P
mappings are general datalog rules) and (ii) a set of domain relations
(also suggested in \cite{SeGiMyBe03}), mapping the objects of one peer
to the objects of another peer. A distributed query answering
algorithm is presented based on synchronous messaging. However, the
algorithm will perform poorly in our restricted framework\footnote{In
  our framework, domain relations correspond to the identity
  relation.}, since when a peer receives a (sub)query, it iterates
through the relevant P2P mappings and for each one of them, sends a
(sub)query to the appropriate peer (waiting for its answer), until
fix-point is reached.  In our case, when a peer receives a (sub)query,
each relevant P2P mapping is considered just once and no iteration
until fix-point is required.

A P2P framework similar to \cite{CalvaneseDBISP2P-2003} is presented
in \cite{Halevy03a}, where query answering according to FOL semantics
is investigated.  Since in general, query answering is undecidable,
the authors present a centralized algorithm (employed in the Piazza
system \cite{HalevyTKDE04}), which however is complete (the algorithm
is always sound), only for the case that polynomial time complexity in
query answering can be achieved. This includes the condition that
inclusion P2P mappings are acyclic.  However, such a condition
severely restricts the modularity of the system. Note that our
algorithm is sound and complete even in the case that there are cycles
in the term dependency path and it always terminates. Thus, our
framework allows placing articulations between peers without further
checks. This is quite important, because the actual interconnections
are not under the control of any actor in the system.

\comment{==In \cite{FranconiKLZ04a,FranconiKLZ04b}, the authors
  consider a framework where each peer is associated with a relational
  database, and P2P mapping rules contain conjunctive queries in both
  the head and the body of the rule (possibly with existential
  variables), each expressed over the alphabet of a single peer. Again
  the semantics of the system is defined based on epistemic logic
  \cite{FranconiKLS03c}.  A distributed query answering algorithm
  (based on asynchronous messaging) is provided where when a peer
  receives a query, the query is processed locally by the peer itself
  using its own data. This first answer is immediately replied back to
  the node which issued the query and sub-queries are propagated to the
  relevant neighbour peers according to the involved mapping rules.
  When a peer receives an new answer, it materializes the view
  represented in the head on the involved mapping rule and propagates
  the answer to the peer that issued the (sub)query.  Answer
  propagation stops when no new answer tuples are coming to the peer
  through any dependency path, that is until fix-point is reached.  Our
  algorithm is also based on asynchronous messaging. However, since it
  considers a limited framework, it is much simpler and for each term
  (sub)query issued to a peer through $\ask$, only one answer is
  returned through $\tell$.  ===}

In \cite{FranconiKLZ04a,FranconiKLZ04b}, the authors consider a
framework where each peer is associated with a relational database,
and P2P mapping rules contain conjunctive queries in both the head and
the body of the rule (possibly with existential variables), each
expressed over the alphabet of a single peer. Again the semantics of
the system is defined based on epistemic logic \cite{FranconiKLS03c}.
In these papers, a peer database update algorithm is provided allowing
for subsequent peer queries to be answered locally without fetching
data from other nodes at query time. The algorithm (which is based on
asynchronous messaging) starts at the peer which sends queries to all
neighbour peers according to the involved mapping rules. When a peer
receives a query, the query is processed locally by the peer itself
using its own data. This first answer is immediately replied back to
the node which issued the query and sub-queries are propagated
similarly to all neighbour peers. When a peer receives an answer, (i)
it stores the answer locally, (ii) it materializes the view
represented in the head on the involved mapping rule, and (ii) it
propagates the result to the peer that issued the (sub)query.  Answer
propagation stops when no new answer tuples are coming to the peer
through any dependency path, that is until fix-point is reached. In our
case, the database update problem for a peer $\ES$ amounts to invoking
$\ES':\query(q)$ for each articulation $q \preceq t$ from $\ES$ to
another peer $\ES'$ and storing the answer locally to $\ES$.  Note
that our query answering algorithm is also based on asynchronous
messaging.  However, since it considers a limited framework, it is
much simpler and no computation until fix-point is required. In
particular, for each term (sub)query issued to a peer through $\ask$,
only one answer is returned through $\tell$.

\paragraph{Relation with information integration}
The literature about information integration distinguishes two main approaches:
the {\em local-as-view} (LAV) and the {\em global-as-view} (GAV) approach (see
\cite{SWWS-2001,LenzeriniPODS02} for a comparison).  In the LAV approach the
contents of the sources are defined as views over the mediator's schema, while
in the GAV approach the mediator's virtual contents are defined as views of the
contents of the sources.  The former approach offers flexibility in representing
the contents of the sources, but query answering is ``hard'' because this
requires answering queries using views (\cite{Duschka97b,Levy01VLDB,Ullman97}).
On the other hand, the GAV approach offers easy query answering (expansion of
queries until getting to source relations), but the addition/deletion of a
source implies updating the mediator view, i.e. the definition of the mediator
relations.  In our case, and if the articulations contain relationships between
single terms, then we have the benefits of both GAV and LAV approaches, \ie~(a)
the query processing simplicity of the GAV approach, as query processing
basically reduces to unfolding the query using the definitions specified in the
mapping, so as to translate the query in terms of accesses (\ie~queries) to the
sources, and (b) the modeling scalability of the LAV approach, \ie~the addition
of a new underlying source does not require changing the previous mappings.  On
the other hand, term-to-query articulations resemble the GAV approach.

\section{Conclusions}
\label{sec:conc}

In this paper, we have presented a model of networked information sources, each
based on a different terminology and endowed with a taxonomy over that
terminology, and logically connected to the other sources via articulations
between respective concepts. The model is very flexible, in that it allows an
articulation to connect a source to several other sources, by letting the terms
in the tail of an articulation to be drawn from several terminologies.

For this kind of systems, we have presented a query evaluation procedure, rooted
on an algorithm for testing the satisfiability of a set of propositional Horn
clauses. Five architectures for implementing this procedure in a distributed
setting have been considered, stemming from two orthogonal criteria: direct
evaluation \emph{vs.} query re-writing, and data allocation. For each of the
resulting five interesting architectures, an implementation has been described,
in terms of processes, communicating asynchronously through several queues. The
design of the architectures and of the underlying communication scheme has
emphasized efficiency and scalability.

The five implementations have been evaluated from a performance point of view,
via simulations. A ranking has resulted from this evaluation, in which the
direct evaluation over a client-server architecture (named CCD) overdoes the
other ones, followed by the architecture in which the taxonomy is centralized
and the queries are re-written before evaluation (named CDR).

Each implementation consists of several processes, each one
implementing a complex algorithm. All these algorithms have been
specified as UML state machines, in order to be simulated. For
reasons of space, it has not been possible to give a complete
account of all this work. In particular, more emphasis has been
given to the implementations that perform best, CCD and
CDR. However, upon request the complete set of UML state machines
is available. Also the code of the simulator is available upon
request.

We believe that the work presented in the paper provides
conclusive knowledge on the addressed problem, and can be used to
derive an engineered system to be put at work on real-word
applications.

\bibliographystyle{plain}
\bibliography{artsources}

\end{document}